\documentstyle[aps,prd,epsfig,latexsym]{revtex}

\def\Tr{{\sf Tr}}
\def\Re{{\sf Re}}
\def\bar{\overline}
\def\hat{\widehat}
\def\tilde{\widetilde}

\def\spose#1{\hbox to 0pt{#1\hss}}
\def\ltapprox{\mathrel{\spose{\lower 3pt\hbox{$\mathchar"218$}}
 \raise 2.0pt\hbox{$\mathchar"13C$}}}
\def\gtapprox{\mathrel{\spose{\lower 3pt\hbox{$\mathchar"218$}}
 \raise 2.0pt\hbox{$\mathchar"13E$}}}
\def\inapprox{\mathrel{\spose{\lower 3pt\hbox{$\mathchar"218$}}
 \raise 2.0pt\hbox{$\mathchar"232$}}}

\begin{document}
%
%
\title{Perturbative matching of the staggered four-fermion operators
for $\epsilon'/\epsilon$ \footnote{LANL preprint number: LA-UR-01-946} }
\author{ Weonjong Lee \footnote{Email: wlee@gita.lanl.gov} \vspace{5mm} }
\address{ Los Alamos National Laboratory, Mail Stop B285,
        Los Alamos, NM 87545, USA 
	\vspace{5mm} }
\date{\today}
\maketitle
\begin{abstract}
Using staggered fermions, we calculate the perturbative corrections to
the bilinear and four-fermion operators that are used in the numerical
study of weak matrix elements for $\epsilon'/\epsilon$.
We present results for one-loop matching coefficients between
continuum operators, calculated in the Naive Dimensional
Regularization (NDR) scheme, and gauge invariant staggered fermion
operators.
Especially, we concentrate on Feynman diagrams of the current-current
insertion type.
We also present results for the tadpole improved operators.
These results, combined with existing results for penguin diagrams,
provide the complete one-loop renormalization of the staggered
four-fermion operators.
Therefore, using our results, it is possible to match a lattice
calculation of $K^0-\bar{K}^0$ mixing and $K\rightarrow\pi\pi$ decays
to the continuum NDR results with all corrections of ${\cal O}(g^2)$
included.
\end{abstract}


\section{Introduction}
The neutral kaon system has been much studied since it was observed
that the strangeness states ($K^0$, $\bar{K}^0$) in the quark model
mix to produce short-lived and long-lived kaons ($K_S$, $K_L$) in
nature.
The discovery of $K_L\rightarrow\pi\pi$ decays revealed that the weak
interaction violates CP symmetry.
In nature, CP asymmetry happens in two ways: indirect and direct CP
violations.
The dominant effect parameterized by $\epsilon$ (the indirect CP
violation) comes from the fact that Nature's neutral kaon mass
eigenstates are not CP-symmetric.
The phenomenon that the weak interaction itself allows
$K_L\rightarrow\pi\pi$ decays directly is referred to as ``direct CP
violation'' and it is parameterized by $\epsilon'$.
The recent results of $\Re(\epsilon'/\epsilon)$ announced by the KTeV
\cite{ktev} and NA48 \cite{na48} collaborations supported the
existence of the direct CP violation.
Even though the Standard Model (SM) of strong and weak interactions
provides a straight-forward qualitative understanding of CP violation
in terms of a single phase ($\delta$), perturbation theory does not
allow a reliable quantitative calculation of the size of CP violation.
In particular, in $K\rightarrow\pi\pi$ decays, the direct CP violation 
is a result of destructive interference from various contributions.
Theoretically, these come from strong interaction effects on hadronic
matrix elements of the effective four-quark operators.
Since the energy involved in these decays is at a scale of around 500
MeV, non-perturbative tools such as lattice QCD must be used to
calculate them.
There have been a number of attempts to calculate $\epsilon'/\epsilon$
on the lattice using clover/Wilson fermions \cite{ape-1,bernard-1},
staggered fermions \cite{kilcup0}, and domain-wall fermions
\cite{blum-1,bob-1,cp-pacs-1}.
In this paper, we study the method using staggered fermions
\cite{tool-kit}.
Recently there has been a numerical study with controlled statistical
errors in calculating the $\Delta I = 1/2$ amplitude, using staggered
fermions \cite{kilcup0}.
However, this numerical study suffers from two major difficulties.
The first is the puzzling crux in calculating $\langle \pi\pi \mid
{\cal O}_i \mid K \rangle$ directly in Euclidean space
\cite{lellouch}.
Therefore, one calculates hadronic matrix elements of $\langle \pi
\mid {\cal O}_i \mid K \rangle$ and $\langle 0 \mid {\cal O}_i \mid K
\rangle$ on the lattice, and reconstructs $\langle \pi\pi \mid {\cal
O}_i \mid K \rangle$ out of these amplitudes, using chiral
perturbation at leading order \cite{soni-1}.
Correspondingly, this method has uncertainties from the neglected
higher orders in chiral perturbation such as final state interactions.
The second difficulty is that the complete set of one-loop matching
formulae for gauge invariant staggered fermion operators was not
available.
Only part of one-loop matching formulae for gauge invariant $B_K$ and
a complete set of one-loop matching relationship for
gauge non-invariant Landau gauge operators were available in
Ref.~\cite{sharpe4,JLQCD1,wlee0}.
Hence, even though the numerical simulation has reached a point to
have statistical control over ``eye'' diagrams \cite{kilcup0}, it was
not quite possible to do the complete one-loop matching of the
gauge invariant lattice results to the continuum hadronic matrix
elements.
In this paper, we present the one-loop matching formulae between the
continuum operators, defined in the NDR scheme, and the lattice gauge
invariant operators of phenomenological interest for the
current-current insertion type (corresponding to the ``eight''
contraction type in Ref.~\cite{kilcup0}).
These results, combined with existing results for penguin diagrams in
Ref.~\cite{sharpe4}, allow matching lattice calculation of
$K^0-\bar{K}^0$ mixing and $K\rightarrow\pi\pi$ decays (including
$\Delta I = 1/2$ amplitudes) to the continuum NDR scheme with all
corrections of ${\cal O}(g^2)$ included.
This paper is organized as follows.
In Sec.~\ref{sec:notation}, we explain our notation for staggered
fermions, bilinears and four-fermion operators on the lattice.
Sec.~\ref{sec:renorm} is devoted to the full description of
renormalization and matching of staggered fermion composite
operators.
We start by presenting Feynman rules, and carry out the
renormalization for bilinear and four-fermion operators on the
lattice.
We explain the tadpole improvement procedure for gauge invariant
staggered fermion operators.
We present the matching formula between lattice and continuum
operators as our final results.
Finite renormalization constants are summarized in Tables
\ref{tab:bi-const-1}--\ref{tab:ff-op-T-S-P-II-2}.
Useful Fierz identities are described in Appendix \ref{app:fierz-1} and
\ref{app:fierz-2}.
We close with some conclusions.

\section{Notations and Terminology} \label{sec:notation}
In this section we will specify our notation for the action, fermion
fields and composite operators on the lattice.
We use the same notation for the staggered fermion action as in
Ref.~\cite{smit1}:
\begin{eqnarray}
S = a^4 \sum_{n} \bigg[ \frac{1}{2a}
\sum_{\mu} \eta_{\mu}(n)
\Big(
\bar{\chi}(n) U_{\mu}(n) \chi(n + \hat{\mu}) -
\bar{\chi}(n + \hat{\mu}) U^{\dagger}_{\mu}(n) \chi(n)
\Big)
+ m \bar{\chi}(n)\chi(n) \bigg] \ ,
\end{eqnarray}
where $ n = (n_1,n_2,n_3,n_4)$ is the lattice coordinate and $
\eta_{\mu}(n) = (-1)^{n_1 + \cdots + n_{\mu-1}} $.

In order to construct the four spin component Dirac field, we adopt
the coordinate space method suggested in Ref.~\cite{klubergstern1}.
In this method, we interpret 16 staggered fermion fields ($\chi$) of
each hypercube in the coordinate space as 4 Dirac spin and 4 flavor
components. The 4 flavor Dirac field $ Q(y) $ is defined as
\begin{eqnarray}
Q^{\alpha i}(y) =
\frac{1}{ N_f \sqrt{N_f}}
\sum_{A} (\gamma_{A})^{\alpha i}
\chi(y_A)
\end{eqnarray}
where $ \alpha $ is the Dirac spin index,
$ i $ is the flavor index and $ N_f $ is the
number of degenerate flavors ($ N_f = 4 $).
Define also
\begin{eqnarray}
y_A \equiv 2 y + A,\ \ \mbox{with }  A \in \{0,1\}^{4}
\end{eqnarray}
and
\begin{eqnarray}
\gamma_A = \gamma^{A_1}_1
\gamma^{A_2}_2 \gamma^{A_3}_3
\gamma^{A_4}_4 \ .
\end{eqnarray}

In order to construct bilinear operators, we use a $\gamma$ matrix
basis for both spin and flavor matrices. Using the Dirac fields $Q$,
we can express a general form of quark bilinear operators:
\begin{eqnarray}
\bar{Q}^{\alpha,i} (y)
(\gamma_S^{\alpha,\beta} \otimes \xi_F^{i,j}) ) Q^{\beta,j} (y)
= \frac{1}{N_f^2} \bar{\chi}(y_A) 
(\overline{\gamma_S \otimes \xi_F })_{AB} \chi(y_B)
\end{eqnarray}
Here, $ \alpha $, $ \beta $ are Dirac spin indices
and $i$, $j$ are flavor indices.
The matrix $(\overline{\gamma_S \otimes \xi_F })$ is defined as
\begin{eqnarray}
( \overline{ \gamma_S \otimes \xi_F})_{AB}
\equiv \frac{1}{N_f}
\mbox{Tr}(\gamma^{\dagger}_A
\gamma_S \gamma_B
\gamma^{\dagger}_F)\ .
\end{eqnarray}
This interpretation is called the {\em coordinate-space method}.

We can translate this coordinate space method directly into
the {\em momentum space method} \cite{kieu1} as follows:
\begin{eqnarray}
\label{k-gs}
(\overline{\overline{ \gamma_S \otimes \xi_F }})_{AB}
=
\sum_{CD} \frac{1}{N_f}(-1)^{A \cdot C}
(\overline{ \gamma_S \otimes \xi_F })_{CD}
\frac{1}{N_f}(-1)^{D \cdot B} \ .
\end{eqnarray}
Note that the momentum-space matrices $ (\overline{\overline{ \gamma_S
\otimes \xi_F }}) $ proposed in Ref.~\cite{smit1} are unitarily
related to the coordinate-space matrices 
$ (\overline{ \gamma_S \otimes \xi_F }) $.

The continuum limit of the staggered fermion action on the lattice
corresponds to QCD with four degenerate flavors ($ N_f = 4 $)
\cite{smit1}. 
In this limit, $ \gamma_S $ represents the Dirac spin matrix and $
\xi_{F} $ belongs to SU(4) flavor symmetry group.
There are numerous choices to transcribe the lattice operator
for a given continuum operator \cite{wlee0,sharpe0,sharpe1}. 
We adopt the following conventional choice of bilinear operator
transcription:
\begin{eqnarray}
\label{bilinear}
{\cal O}_{SF}(y) = \frac{1}{N_f}
\sum_{AB}\bar{\chi}(y_A)
( \overline{\gamma_S \otimes \xi_F } )_{AB}
\chi(y_B) \ {\cal U}(y_A,y_B) \;.
\end{eqnarray}
Here, $ {\cal U}(y_A,y_B) $ is a product of gauge links that makes $
{\cal O}_{SF} $ gauge invariant \cite{wlee0,sheard0,sharpe1,sheard1}.
The link matrices $ {\cal U}(y_A,y_B) $ are constructed by averaging
over all of the shortest paths between $ y_A $ and $ y_B $
\cite{wlee0,sheard0,sheard1}, such that $ {\cal O}_{SF} $ is as
symmetric as possible:
\begin{eqnarray}
\label{ulink}
{\cal U}(y_A,y_B) & = & \frac{1}{4!}
\sum_{P} U(y_A, y_A+\Delta_{P1})\cdots
\nonumber \\ & &
U(y_A+\Delta_{P1}+\Delta_{P2}+\Delta_{P3}\ ,\ y_B) \ ,
\end{eqnarray}
where
$ P $ is an element of the permutation group (1234) and
\begin{eqnarray}
\label{delta}
\Delta_{\mu} = (B_{\mu}-A_{\mu}) \hat{\mu} \ .
\end{eqnarray}
There are two kinds of four-fermion operators which have different
color contraction structure.  The general form of color two-trace
operators is
\begin{equation}
[\bar{\chi}_{a}^{f_1} (y_A)
( \overline{\gamma_S \otimes \xi_F } )_{AB}
\chi_{b}^{f_2} (y_B)]
[\bar{\chi}_{c}^{f_3} (y_C)
( \overline{\gamma_{S'} \otimes \xi_{F'} } )_{CD}
\chi_{d}^{f_4} (y_D)]
\ {\cal U}_{ab}(y_A,y_B)
\ {\cal U}_{cd}(y_C,y_D)
\end{equation}
and the general form of color one-trace operators is
\begin{equation}
[\bar{\chi}_{a}^{f_1} (y_A)
( \overline{\gamma_S \otimes \xi_F } )_{AB}
\chi_{b}^{f_2} (y_B)]
[\bar{\chi}_{c}^{f_3} (y_C)
( \overline{\gamma_{S'} \otimes \xi_{F'} } )_{CD}
\chi_{d}^{f_4} (y_D)]
\ {\cal U}_{ad}(y_A,y_D)
\ {\cal U}_{cb}(y_C,y_B)
\end{equation}
where $f_1$, $f_2$, $f_3$ and $f_4$
label the continuum
flavor (for example, u, d ,s, $ \cdots $).
Note that we have introduced flavor in two ways: first, the 4
degenerate flavors hidden in the hypercubic index $A$ of the staggered
fermion field $\chi^f (y_A)$, and second, the continuum flavors
labeled by the index $f$.
For later convenience, let us introduce a compact notation.  We can
represent the bilinear and four-fermion operators in terms of $ S $
(scalar), $ V $ (vector) , $ T $ (tensor), $ A $ (axial) and $ P $
(pseudo-scalar). For example, the bilinears can be expressed as
\begin{eqnarray}
[V_\mu \times S]
& \equiv & \frac{1}{N_f} \sum_{A,B}
[\bar{\chi}(y_A)
(\overline{ \gamma_{\mu} \otimes I } )_{AB}
\chi(y_B)] {\cal U}(y_A,y_B)
\\ \hspace*{0mm}
[A_\mu \times P]
& \equiv  & \frac{1}{N_f} \sum_{A,B}
[\bar{\chi}(y_A)
(\overline{ \gamma_{\mu5} \otimes \xi_5 } )_{AB}
\chi(y_B) ] {\cal U}(y_A,y_B)
\end{eqnarray}
Here, the continuum flavor indices are dropped out.
For the four-fermion operators, we use the same notation as the
bilinears but need to distinguish between color one trace and color
two trace operators.
This notation is exemplified by the following:
\begin{eqnarray}
[S \times P] [P \times S]_{II}
& \equiv & \frac{1}{N_f^2} \sum_{A,B,C,D}
[\bar{\chi}(y_A) (\overline{ I \otimes \xi_5 } )_{AB} \chi(y_B)] 
[\bar{\chi}(y_C) (\overline{ \gamma_5 \otimes I } )_{CD} \chi(y_D)] 
\nonumber \\ & & \hspace*{25mm}
\cdot {\cal U}(y_A,y_B) {\cal U}(y_C,y_D)
\\ \hspace*{0mm}
[A_\mu \times P] [V_\mu \times S]_{I}
& \equiv  & \frac{1}{N_f^2} \sum_{A,B,C,D}
[\bar{\chi}(y_A) (\overline{ \gamma_{\mu5} \otimes \xi_5 } )_{AB} \chi(y_B) ] 
[\bar{\chi}(y_C) (\overline{ \gamma_{\mu} \otimes I } )_{CD} \chi(y_D) ] 
\nonumber \\ & & \hspace*{25mm}
\cdot {\cal U}(y_A,y_D) {\cal U}(y_C,y_B)
\end{eqnarray}
Here note that the sub-indices $I$, $II$ represent the color one trace
and color two trace operators respectively.

There are two completely independent methods to transcribe the lattice
operators using Fierz transformation: one spin trace formalism and two
spin trace formalism \cite{wlee0}.
In this paper, we choose two spin trace formalism to construct the
lattice operators and it is also adopted for our numerical study on
$\epsilon'/\epsilon$.

\section{Renormalization and Matching} \label{sec:renorm}
In this section, we will review the renormalization of the continuum
operators and explain the matching procedure between the continuum and
the lattice operators at one loop level.
We use the formulation of
Ref.~\cite{sheard0,sheard1,sharpe3,JLQCD1,sharpe4,wlee0}.

\subsection{Feynman rules} \label{subsec:feynman}
We use the same Feynman rules as in Ref.~\cite{wlee0}.
There was a typo in the Feynman rule for the gluon propagator in
Eq.~(78) of Ref.~\cite{wlee0}.
Therefore, we provide the correct Feynman rule here.
\begin{eqnarray}
D_{\mu \nu}^{IJ}(k) =
\frac{ \delta_{IJ}\delta_{\mu\nu} }
{ \sum_{\beta} \frac{4}{a^2} \sin^{2}( \frac{1}{2} ak_{\beta} ) }
-
(1-\alpha)
\frac{ \delta_{IJ} \frac{4}{a^2}
\sin( \frac{1}{2} ak_{\mu} ) \sin( \frac{1}{2} ak_{\nu} )}
{ [ \sum_{\beta} \frac{4}{a^2} \sin^{2}( \frac{1}{2} ak_{\beta} ) ]^2 } \ ,
\end{eqnarray}
where $\alpha$ is a gauge fixing parameter in the general covariant
gauge and superscripts $I,J$ represent the color indices in the
adjoint representation of SU(3).

All the Feynman diagrams are calculated at zero quark mass, which
induces infrared divergences.
We regulate the infrared divergences by adding a mass term to the
gluon propagators both on the lattice and in the continuum.
This dependence on the gluon mass cancels out, when we match the
lattice and continuum renormalized operators.

\subsection{Bilinear Operators} \label{subsec:bi_op}
Let us consider a bilinear operator, ${\cal O}_{SF}$ with spin $S$ and
flavor $F$.
%
%
In the $\overline{MS}$ scheme, there are three distinct sets of
commutation rules regarding $ \gamma_5 $: Naive Dimensional
Regularization (NDR), Dimensional Reduction (DRED, DR$\rm
\overline{EZ}$), and t'Hooft-Veltman prescription (HV)
\cite{buras0,bernard0,sharpe3}.
Here, we choose NDR, the most commonly used scheme.
At one loop level, we can write a general form of the renormalized
continuum operators as
\begin{eqnarray}
\label{bilinearc}
\langle {\cal O}^{Cont(1)}_{SF} \rangle = \sum_{S'F'}
( \delta_{SS'} \delta_{FF'} +
\frac{g^2}{(4\pi)^2} Z^{Cont}_{SF;S'F'} )
\langle {\cal O}^{Cont(0)}_{S'F'} \rangle \ .
\end{eqnarray}
where $\langle \cdots \rangle$ represent amputated Green's
functions between external quark states.
Here, the superscript $ (i) $
($ i \in \{0,1,2,\cdots\}$)
represents the number of loops and
\begin{eqnarray}
\label{eq:bi-cont-Z}
Z^{Cont}_{SF;S'F'} \equiv
\delta_{SS'}\delta_{FF'} \Big[\Gamma_S \log(\frac{\mu}{\kappa})
+ C^{Cont}_{S}
+ R_{S} \Big] \ .
\end{eqnarray}
The $ \mu $ is the renormalization scale and $ \kappa $ is the gluon
mass, which regulates the infra-red.  
$ \Gamma_S $ is the anomalous dimension matrix:
\begin{eqnarray}
\label{eq:anomalous1}
\Gamma_S & = & \frac{8}{3}(\sigma_S -1)
\\
\sigma_S \gamma_S & = & \frac{1}{4}\sum_{\mu,\nu}
\gamma_{\mu}\gamma_{\nu}\gamma_{S}
\gamma_{\nu} \gamma_{\mu}\ .
\end{eqnarray}
Note that the vector and axial currents have vanishing anomalous
dimensions.
$ C^{Cont}_{S} $ is a finite constant term which depends on the
regularization and renormalization scheme.
For the NDR scheme \cite{buras0}, 
\begin{eqnarray}
\sigma_S & = & (4,1,0,1,4) \ , \\ 
C^{Cont}_{S} & = &
(\frac{10}{3}, 0 ,\frac{2}{3}, 0, \frac{10}{3}) \mbox{ for } \gamma_S
= (I,\gamma_{\mu}, \sigma_{\mu\nu},\gamma_{\mu5},\gamma_{5} ) \mbox{
respectively.} 
\label{eq:bi-cont-const} 
\end{eqnarray}
The $ R_{S} $ terms contain logarithmic dependence on the external
quark momentum due to anomalous dimension.
They are universal ({\em i.e.} independent of the regularization and
renormalization scheme).
%

%
%
On the lattice, there are eight Feynman diagrams (Figure 2 in
Ref.~\cite{wlee0}) contributing to bilinear operators at one loop
level.
Note that only half of the self-energy diagrams (Figure 2 (e), (f),
(g) and (h) in Ref.~\cite{wlee0}) contribute since the other half are
absorbed in the wave-function renormalization of the external quark
fields.
The explicit analytic results of these diagrams are given in
Appendix A of Ref.~\cite{JLQCD1}.
The renormalized lattice bilinear operators at one loop level can be
expressed in terms of tree level lattice operators as
\begin{eqnarray}
\label{eq:bilinearl}
\langle {\cal O}^{Latt(1)}_{SF} \rangle
= \sum_{S'F'}
(\delta_{SS'} \delta_{FF'} +
\frac{g^2}{(4\pi)^2} Z^{Latt}_{SF;S'F'} ) \ 
\langle {\cal O}^{Latt(0)}_{S'F'} \rangle \ ,
\end{eqnarray}
where
\begin{eqnarray}
Z^{Latt}_{SF;S'F'} \equiv
-\delta_{SS'}\delta_{FF'} \Gamma_S \log(a{\kappa})
+ C^{Latt}_{SF;S'F'}
+ \delta_{SS'}\delta_{FF'} R_{S} \ .
\label{eq:latt-Z}
\end{eqnarray}
Here, $ \Gamma_S $ is the same as in the continuum (see
Eq.~(\ref{eq:anomalous1})) since the anomalous dimension at one loop
is independent of the regularization and renormalization schemes.
$ C^{Latt}_{SF;S'F'} $ is the finite constant term unique to the
lattice regularization scheme.
Numerical values of $C^{Latt}_{SF;S'F'}$ are summarized in Tables
\ref{tab:bi-const-1} and \ref{tab:bi-const-2}.
%
%
In Ref.~\cite{sharpe4,wlee0}, it is proven that the $U_A(1)$ symmetry
of the staggered fermion action insures that for any spin-flavor
structure the two bilinear operators $ \bar{\chi}_s (\overline{
\gamma_{S} \otimes \xi_{F} }) \chi_d $ and $ \bar{\chi}_s (\overline{
\gamma_{S5} \otimes \xi_{F5} }) \chi_d $ are renormalized identically
regardless of their distance.
Hence, in Tables \ref{tab:bi-const-1} and \ref{tab:bi-const-2}, we
provide the $ C^{Latt}_{SF;S'F'} $ values for only one of these two
bilinear operators.
Our values of the $C^{Latt}_{SF;S'F'}$ constants are related to the $
c_{ij} $ ($i=SF$ and $j=S'F'$) in Tables 6 and 7 of
Ref.~\cite{sharpe3} as follows:
\begin{eqnarray}
& & c_{ij} = \frac{1}{C_F} \bigg[ \delta_{SS'}\delta_{FF'}
\Big( C^{Cont,\rm DR\overline{EZ}}_S + \Gamma_S \log(\pi) \Big) 
- C^{Latt}_{SF;S'F'} \bigg]
\\
& & C^{Latt}_{SF;S'F'}
= - C_F \bigg[ c_{ij} - \delta_{ij} \Big(  
(\sigma_S - 1) (2 \log(\pi) + 1) + t_S \Big) \bigg] \, ,
\end{eqnarray}
where 
\begin{eqnarray}
C_F = \frac{N^2_c - 1}{2N_c} = \frac{4}{3} \, .
\end{eqnarray}
Here, note that $ c_{ij} $ is calculated for the DR$\rm\overline{EZ}$
scheme and
\begin{eqnarray}
C^{Cont,\rm DR\overline{EZ}}_{S} & = &
(\frac{14}{3}, 0 ,-\frac{2}{3}, 0, \frac{14}{3}) \mbox{ for } \gamma_S
= (I,\gamma_{\mu}, \sigma_{\mu\nu},\gamma_{\mu5},\gamma_{5} ) \mbox{
respectively.}
\end{eqnarray}
The values of $t_S$ for DR$\rm\overline{EZ}$ are the same as in
Ref.~\cite{sharpe3}.
\begin{eqnarray}
t_S & = &
(\frac{1}{2}, 0 ,\frac{1}{2}, 0, \frac{1}{2}) \mbox{ for } \gamma_S
= (I,\gamma_{\mu}, \sigma_{\mu\nu},\gamma_{\mu5},\gamma_{5} ) \mbox{
respectively.}
\end{eqnarray}
The $C^{Latt}_{SF;S'F'}$ values in this paper are in agreement with
those of Ref.~\cite{sharpe3} within the statistical uncertainty of
finite integrals obtained using the Monte Carlo integration method.
However, our results (and those of Ref.~\cite{sharpe3}) for the scalar
bilinears $[S \times X]$ ($X \in \{S,V,T,A,P\}$) and the mixing of the
bilinear $[V_\mu \times V_\nu]$ with $[V_\mu \times V_\mu]$ are in
disagreement with those in Ref.~\cite{JLQCD1}, while the rest of our
results agree.
%
%
%
%
%
%
We have presented the one-loop renormalized operators in terms of tree
level operators both in the continuum and on the lattice.
Now we need to relate the lattice and continuum operators.
The matching condition is based upon the observation that at tree
level, the continuum operators are, by construction, the same as
lattice operators in the limit of the zero lattice spacing ($a
\rightarrow 0 $).
In other words, at tree level as $ a \rightarrow 0 $, the lattice
operators are related to the continuum operators as follows:
\begin{eqnarray}
\label{connect}
{\cal O}^{Latt(0)}_{SF} = {\cal O}^{Cont(0)}_{SF} +
{\cal O}(a) \ .
\end{eqnarray}
Here, ${\cal O}(a)$ represents terms of order $ a $ or higher, which
are supposed to vanish as $ a \rightarrow 0 $.
At one loop (and higher) order, the lattice and the continuum operators
will differ. 
A carefully constructed mixture of lattice operators is needed to
reproduce the desired continuum operator.
From Eqs.~(\ref{bilinearc}), (\ref{eq:bilinearl}), (\ref{connect}), we
can connect the lattice operators with the continuum operators at one
loop level via
\begin{eqnarray}
{\cal O}^{Cont(1)}_{SF} & = &
\sum_{S'F'} \left[ \delta_{SS'}\delta_{FF'} +
\frac{ g^2 }{ (4\pi)^2 }
\left( Z^{Cont}_{SF;S'F'} - Z^{Latt}_{SF;S'F'} \right)
\right] {\cal O}^{Latt(1)}_{S'F'}
\nonumber \\
& = & \sum_{S'F'}
\left[ \delta_{SS'}\delta_{FF'}
+ \frac{ g^2 }{ (4\pi)^2 }
\delta_{SS'} \delta_{FF'} \Gamma_S \ln(\mu a) +
\right.
\nonumber \\
& & \left. \frac{ g^2 }{ (4\pi)^2 }
\left(
\delta_{SS'} \delta_{FF'} C^{Cont}_S
-C^{Latt}_{SF;S'F'}
\right) \right]
{\cal O}^{Latt(1)}_{S'F'}
\label{eq:match-bi}
\end{eqnarray}
This is the matching formula between the lattice and continuum
bilinear operators.

%
%
Lepage, Mackenzie, and Parisi observed that the tadpole diagrams cause
the large difference between the bare gauge coupling $ g_{0}(a) $ and
the renormalized coupling $ g_{\overline{MS}}( \mu = 1/a ) $
\cite{parisi,lepage}.
They proposed a mean field method for removing the dominant effect of
tadpole diagrams.
This method of tadpole improvement has been applied to quark mass
\cite{wlee1} and composite operator renormalization
\cite{sharpe3,sharpe4,JLQCD1,wlee0} extensively.
Here, let us explain the tadpole improvement procedure for the
bilinear operators.
The first step in this procedure is the rescaling of the staggered
fermion fields and gauge link fields \cite{wlee0,wlee1}.
\begin{eqnarray}
\chi \rightarrow \psi =\sqrt{u_0} \chi
\nonumber \\
\bar{\chi} \rightarrow \bar{\psi} = \sqrt{u_0}
\bar{\chi}
\nonumber \\
U_{\mu} \rightarrow \tilde{U}_\mu = \frac{ U_{\mu} }{ u_0} \ .
\label{eq:rescale}
\end{eqnarray}
A gauge invariant choice of the mean field scaling factor $u_0$ is 
\cite{lepage,karsch}
\begin{eqnarray}
u_0 = \bigg[ \frac{1}{3} \Re \langle  \Tr U_\Box \rangle \bigg]^{1/4}
    = 1 - \frac{1}{12} g^2 + O(g^4)  \, .
\end{eqnarray}
Under this rescaling, the bilinear operators transform as
\begin{eqnarray}
{\cal O}^{latt(1)}_{SF} \rightarrow  \tilde{\cal O} ^{latt(1)}_{SF}
&=& \frac{1}{N_f} \ 
\bar\psi (y_A) (\overline{\gamma_S \otimes \xi_F})_{AB} \psi(y_B)
\tilde{\cal U}_{AB} 
\nonumber \\
&=&  
u_0^{1-\Delta_{SF}} \cdot \frac{1}{N_f} \  
\bar\chi (y_A) (\overline{\gamma_S \otimes \xi_F})_{AB} \chi(y_B)
{\cal U}_{AB}
\nonumber \\
&=& u_0^{1-\Delta_{SF}} \cdot {\cal O}^{latt(1)}_{SF} \,.
\end{eqnarray}
Here, $ \Delta_{SF} = \sum_\mu \mid S_\mu - F_\mu \mid$ corresponds to
the distance between the quark and antiquark fields of the bilinear
operator.
The point is that the tadpole improved operators $\tilde{\cal
O}^{latt(1)}_{SF}$ are better representations of the continuum
operators ${\cal O}^{Cont(1)}_{SF}$.
Now we can rewrite Eqs.~(\ref{eq:bilinearl},\ref{eq:latt-Z}), the one
loop relationship on the lattice, in terms of the tadpole improved
operators.
\begin{eqnarray}
\label{eq:latt-bi-TI}
\langle \tilde{\cal O}^{Latt(1)}_{SF} \rangle
&=& \sum_{S'F'}
(\delta_{SS'} \delta_{FF'} +
\frac{\tilde{g}^2}{(4\pi)^2} Z^{Latt}_{SF;S'F'} )
\cdot u_0^{1 - \Delta_{SF}} \cdot
\langle {\cal O}^{Latt(0)}_{S'F'} \rangle
\nonumber \\
&=& \sum_{S'F'}
(\delta_{SS'} \delta_{FF'} +
\frac{\tilde{g}^2}{(4\pi)^2} \tilde{Z}^{Latt}_{SF;S'F'} )
\langle {\cal O}^{Latt(0)}_{S'F'} \rangle \ ,
\end{eqnarray}
where
\begin{eqnarray}
\tilde{Z}^{Latt}_{SF;S'F'} &\equiv&
-\delta_{SS'}\delta_{FF'} \Gamma_S \log(a{\kappa})
+ \tilde{C}^{Latt}_{SF;S'F'}
+ \delta_{SS'}\delta_{FF'} R_{S} \, ,
\label{eq:latt-Z-TI}
\\
\tilde{C}^{Latt}_{SF;S'F'} &\equiv& 
C^{Latt}_{SF;S'F'} - (1 - \Delta_{SF}) \frac{ (4\pi)^2 } { 12 }
\delta_{SS'}\delta_{FF'}  \, .
\end{eqnarray}
Here, note that the gauge coupling constant is also tadpole improved:
$g^2\rightarrow \tilde{g}^2$. Regarding the tadpole improvement of
gauge coupling constants, refer to Refs.~\cite{lepage,wlee0,wlee1}
for details.
Correspondingly, we can rewrite Eq.~(\ref{eq:match-bi}), the relationship
between the continuum and lattice operators, as follows:
\begin{eqnarray}
{\cal O}^{Cont(1)}_{SF} & = &
\sum_{S'F'} \left[ \delta_{SS'}\delta_{FF'} +
\frac{ \tilde{g}^2 }{ (4\pi)^2 }
\left( Z^{Cont}_{SF;S'F'} - \tilde{Z}^{Latt}_{SF;S'F'} \right)
\right] \tilde{\cal O}^{Latt(1)}_{S'F'}
\nonumber \\
& = & \sum_{S'F'}
\left[ \delta_{SS'}\delta_{FF'}
+ \frac{ \tilde{g}^2 }{ (4\pi)^2 }
\delta_{SS'} \delta_{FF'} \Gamma_S \ln(\mu a) +
\right.
\nonumber \\
& & \left. \frac{ \tilde{g}^2 }{ (4\pi)^2 }
\left(
\delta_{SS'} \delta_{FF'} C^{Cont}_S
-\tilde{C}^{Latt}_{SF;S'F'}
\right) \right]
\tilde{\cal O}^{Latt(1)}_{S'F'} \, .
\end{eqnarray}
This is the tadpole-improved matching formula for the bilinear
operators.

\subsection{Four Fermion Operators} \label{subsec:ff-op}
%
%
Let us consider a general form of the staggered four-fermion operators
on the lattice:
\begin{eqnarray*}
[ \bar{\chi} (\overline{ \gamma_S \otimes \xi_F }) \chi ]
[ \bar{\chi} (\overline{ \gamma_{S'} \otimes \xi_{F'} }) \chi ]
\end{eqnarray*}
There are $16^4 = 65536$ four-fermion operators. 
Hence, the mixing matrices are naively $65536 \times 65536$.
However, we are interested in only a subset of these huge mixing
matrices.
In this paper, we focus on operators of phenomenological interest,
especially relevant to the direct CP violation, $\epsilon'/\epsilon$.
%

%
%
On the lattice, gauge non-invariant four-fermion operators such as
Landau-gauge operators mix with gauge non-invariant lower dimension
bilinear operators.
This requires additional non-perturbative subtractions and it is
significantly more difficult to extract the divergent mixing
coefficients in a completely non-perturbative way.
For this reason, we are not interested in gauge non-invariant
operators.
Therefore, we study gauge invariant operators in this paper, since the
gauge invariance protects operators from mixing with lower dimension
gauge non-invariant operators \cite{sharpe4}.
%

%
%
Let us first classify the Feynman diagrams into two categories.
One is the current-current insertion type and the other is the penguin
insertion type\cite{buras1,sharpe4}.
In the case of the penguin diagrams, the perturbative corrections and
the matching formula for the gauge invariant operators are given at
one loop in Ref.~\cite{sharpe4}.
However, for the current-current diagrams, the perturbative matching
formulae at one loop level are not available for the complete set of
the gauge invariant staggered fermion operators of phenomenological
interest ($\epsilon'/\epsilon$).
Therefore, in this paper, we calculate perturbatively the
current-current diagrams of these operators and obtain the
corresponding matching formulae at one loop.
%

%
%
We begin by reviewing the renormalization of the four-fermion
operators for the current-current insertion type in the continuum.
The continuum operators of phenomenological interest can be
expressed in terms of the practical basis introduced in
Ref.~\cite{sharpe5}.
\begin{eqnarray}
{\cal O}_1 &=& (\bar{\psi}_1 \gamma_\mu L \psi_2) 
		(\bar{\psi}_3 \gamma_\mu L \psi_4) 
\nonumber \\ 
{\cal O}_2 &=& (\bar{\psi}_1 \gamma_\mu L \psi_2) 
		(\bar{\psi}_3 \gamma_\mu R \psi_4) 
\nonumber \\
{\cal O}_3 &=& -2 (\bar{\psi}_1 L \psi_2) 
		(\bar{\psi}_3 R \psi_4)
\label{eq:cont-basis}
\end{eqnarray}
where $ L = (1-\gamma_5)$ and $R = (1 + \gamma_5)$.
There are two more operators included in the practical basis for
completeness in Ref.~\cite{sharpe5}.
\begin{eqnarray}
{\cal O}_4 &=&  (\bar{\psi}_1 L \psi_2) 
		(\bar{\psi}_3 L \psi_4)
\nonumber \\
{\cal O}_5 &=&  - \frac{1}{8} \Bigl( 1 + \frac{3}{4} \varepsilon \Bigr)
		\sum_{\mu,\nu}	
		(\bar{\psi}_1 \gamma_\mu \gamma_\nu L \psi_2) 
		(\bar{\psi}_3 \gamma_\nu \gamma_\mu L \psi_4)
\label{eq:cont-basis-2}
\end{eqnarray}
The odd parity parts of these operators are not important for the
following discussion, but they are included for generality.
As in the bilinear operators, we choose NDR as our continuum
renormalization scheme.
In this scheme, there are many ways to extend the definition of
four-fermion operators to $n= 4- 2\varepsilon$ dimensions, which
corresponds to different choices of evanescent operators.
Here, we adopt the convention of Ref.~\cite{sharpe5}.
In this convention, the perturbative corrections of four-fermion
operators are derived from those of bilinear operators by imposing the
Fierz symmetry and by loosening the charge conjugation.
Using this convention, we can express a general form of the
renormalized continuum four-fermion operators as
\begin{eqnarray}
\vec{\cal O}_i &\equiv& \left( 
			\begin{array}{c} 
			{\cal O}_{i,I} \\ {\cal O}_{i,II}
			\end{array} \right) 
\\
\langle \vec{\cal O}^{Cont(1)}_i \rangle &=& 
\langle \vec{\cal O}^{Cont(0)}_i \rangle + 
\frac{g^2}{(4\pi)^2} \cdot \hat{Z}^{Cont}_{ij} 
\cdot \langle \vec{\cal O}^{Cont(0)}_j \rangle
\label{eq:ff-cont-Z}
\end{eqnarray}
Here, note that $\hat{Z}^{Cont}_{ij} = \delta_{ij} \hat{Z}^{Cont}_i$
for $ i = \{ 1,2,3\}$, and $\hat{Z}^{Cont}_i$ is a $ 2 \times 2 $
matrix defined as
\begin{eqnarray}
\hat{Z}^{Cont}_1 &=& z_{V+A} \hat{C}_a + z_{V+A} \hat{C}_b 
		- (z_{S+P} + 6) \hat{C}_c
\nonumber \\
\hat{Z}^{Cont}_2 &=& z_{V+A} \hat{C}_a + z_{S+P} \hat{C}_b 
		- (z_{V+A} + 6) \hat{C}_c
\nonumber \\
\hat{Z}^{Cont}_3 &=& z_{S+P} \hat{C}_a + z_{V+A} \hat{C}_b 
		- (z_{V+A} + 6) \hat{C}_c
\end{eqnarray}
The $\vec{\cal O}_4$ and $\vec{\cal O}_5$ operators mix with each other.
\begin{eqnarray}
\hat{Z}^{Cont}_{44} &=& z_{S+P} \hat{C}_a 
			+ 2 z_{T} \hat{C}_b 
			- \frac{1}{2} ( z_{S+P} + 2 z_{T} + 4 ) \hat{C}_c
\nonumber \\
\hat{Z}^{Cont}_{45} &=& (-1) \hat{C}_a 
			+ \frac{1}{2} ( 2 z_{T} - z_{S+P} ) \hat{C}_b 
			- \frac{1}{2} ( z_{S+P} - 2 z_{T} + 2 ) \hat{C}_c
\nonumber \\
\hat{Z}^{Cont}_{54} &=& \frac{1}{2} ( 2 z_{T} - z_{S+P} ) \hat{C}_a 
			+ (-1) \hat{C}_b 
			- \frac{1}{2} ( z_{S+P} - 2 z_{T} + 2 ) \hat{C}_c
\nonumber \\
\hat{Z}^{Cont}_{55} &=& ( 2 z_{T} ) \hat{C}_a 
			+ z_{S+P} \hat{C}_b 
			- \frac{1}{2} ( z_{S+P} + 2 z_{T} + 4 ) \hat{C}_c
\end{eqnarray}
Here, the color mixing matrices $ \hat{C} $ are
\begin{eqnarray}
\hat{C}_a = \frac{1}{6} \left( \begin{array}{cc}
			-1 & 3 \\ 0 & 8 
			\end{array} \right) \, , \quad
\hat{C}_b = \frac{1}{6} \left( \begin{array}{cc}
			8 & 0 \\ 3 & -1 
			\end{array} \right) \, , \quad
\hat{C}_c = \frac{1}{6} \left( \begin{array}{cc}
			-1 & 3 \\ 3 & -1 
			\end{array} \right) \, .
\end{eqnarray}
The renormalization coefficients $z_{V+A}$, $z_{S+P}$ and $z_{T}$ can
be expressed in terms of the bilinear corrections given in Eq.~
(\ref{eq:bi-cont-Z}) as follows:
\begin{eqnarray}
& & z_{V+A} \equiv z_V + z_A
\\ & & 
z_{S+P} \equiv z_S + z_P
\\ & & 
z_X \equiv \frac{1}{C_F} \Big[\Gamma_X \log(\frac{\mu}{\kappa}) +
C^{Cont}_{X} + R_{X} \Big] \,.
\end{eqnarray}
Here, $ X \in \{S,V,T,A,P\}$ represents the spin structure of bilinear
operators and $\Gamma_X$, $C^{Cont}_{X}$ are given in
Eqs.~(\ref{eq:anomalous1}) and (\ref{eq:bi-cont-const}) respectively.
Note that $R_X$ terms will cancel out when we match between the
lattice and continuum operators.
For the NDR and DR$\rm \overline{EZ}$ schemes, $z_V=0$ and $z_A=0$, and
so $z_{V+A}=0$. This simplifies the $\hat{Z}^{Cont}_i$ further:
\begin{eqnarray}
\hat{Z}^{Cont}_1 &=& - (z_{S+P} + 6) \hat{C}_c
\nonumber \\
\hat{Z}^{Cont}_2 &=& z_{S+P} \hat{C}_b - 6 \hat{C}_c
\nonumber \\
\hat{Z}^{Cont}_3 &=& z_{S+P} \hat{C}_a - 6 \hat{C}_c \, .
\end{eqnarray}
As a summary, we present the results of the renormalization of the
continuum four-fermion operators in the practical basis as follows:
\begin{eqnarray}
\langle \vec{\cal O}^{Cont(1)}_i \rangle &=& 
\langle \vec{\cal O}^{Cont(0)}_i \rangle + \frac{g^2}{(4\pi)^2} \Big[
	\hat{\Gamma}_{ij} \log(\frac{\mu}{\kappa}) + 
	\hat{C}^{Cont}_{ij} + \hat{R}_{ij} \Big]
	\langle \vec{\cal O}^{Cont(0)}_{j} \rangle
\label{eq:ff-cont-Z-2}
\end{eqnarray}
Here, note that $\hat{\Gamma}_{ij} = \delta_{ij} \hat{\Gamma}_i$ and
$\hat{C}^{Cont}_{ij} =  \delta_{ij} \hat{C}^{Cont}_i$
for $ i \in \{1,2,3\}$.
The anomalous dimension matrices are
\begin{eqnarray}
\hat{\Gamma}_1 &=& \hat{\Gamma}_{V+A} = 
	\left( \begin{array}{cc}
	 +2 & -6 \\ -6 & +2 
	\end{array} \right)
\nonumber \\
\hat{\Gamma}_2 &=& \hat{\Gamma}_{V-A} = 
	\left( \begin{array}{cc}
	 +16 & 0 \\ +6 & -2 
	\end{array} \right)
\nonumber \\
\hat{\Gamma}_3 &=& \hat{\Gamma}_{2[P-S]} = 
	\left( \begin{array}{cc} 
	-2 & +6 \\ 0 & +16 
	\end{array} \right)
\nonumber 
\end{eqnarray}
\begin{eqnarray}
\hat{\Gamma}_{44} &=& + \frac{4}{3} \cdot
	\left( \begin{array}{cc} 
	-5 & +3 \\ -3 & +13
	\end{array} \right)
\nonumber \\
\hat{\Gamma}_{45} &=& - \frac{4}{3} \cdot
	\left( \begin{array}{cc} 
	+7 & +3 \\ +6 & -2
	\end{array} \right)
\nonumber \\
\hat{\Gamma}_{54} &=& - \frac{4}{3} \cdot
	\left( \begin{array}{cc} 
	 -2 & +6 \\ +3 & +7
	\end{array} \right)
\nonumber \\
\hat{\Gamma}_{55} &=& + \frac{4}{3} \cdot
	\left( \begin{array}{cc} 
	+13 & -3 \\ +3 & -5 
	\end{array} \right) \,.
\label{eq:ff-anomalous}
\end{eqnarray}
These are universal and independent of the renormalization scheme.
For the NDR scheme, the finite constant matrices are
\begin{eqnarray}
\hat{C}^{Cont}_1 &=& \hat{C}^{Cont}_{V+A} 
		= \frac{11}{12} \hat{\Gamma}_1
		= 	\left( \begin{array}{cc} 
			 +11/6 & -11/2 \\
			 -11/2 & +11/6 
			\end{array} \right)
\nonumber \\
\hat{C}^{Cont}_2 &=& \hat{C}^{Cont}_{V-A} = 
	\left( \begin{array}{cc} 
	 +23/3 & -3 \\ 
	 -1/2 & +1/6 
	\end{array} \right)
\nonumber \\
\hat{C}^{Cont}_3 &=& \hat{C}^{Cont}_{2[P-S]} = 
	\left( \begin{array}{cc} 
	 +1/6 & -1/2 \\
	 -3  & +23/3
	\end{array} \right)
\nonumber
\end{eqnarray}
\begin{eqnarray}
\hat{C}^{Cont}_{44} &=& + \frac{2}{3}
	\left( \begin{array}{cc} 
	+2 & 0 \\
	-3 & +11 
	\end{array} \right)
\nonumber \\
\hat{C}^{Cont}_{45} &=& - \frac{1}{2}
	\left( \begin{array}{cc} 
	+4 & +4 \\ 
	+5 & +1 
	\end{array} \right)
\nonumber \\
\hat{C}^{Cont}_{54} &=& - \frac{1}{2}
	\left( \begin{array}{cc} 
	 +1 & +5 \\
	 +4 & +4 
	\end{array} \right)
\nonumber \\
\hat{C}^{Cont}_{55} &=& + \frac{2}{3}
	\left( \begin{array}{cc} 
	 +11 & -3 \\
	   0 & +2 
	\end{array} \right)
\end{eqnarray}
For further details of other continuum renormalization schemes, refer
to Ref.~\cite{sharpe5} and references in it.
%
%
%

%
%
This completes our review of the renormalization of the the
four-fermion operators of the current-current insertion type in the
continuum.
We now describe renormalization of staggered four-fermion
operators of the same type on the lattice.
For the purpose of matching, it is convenient to work with the
practical basis as in the continuum.
Hence, we choose the same practical basis on the lattice.
Using the compact notation introduced in Section \ref{sec:notation} in
the two spin trace formalism \cite{wlee0}, we can express the
practical basis as follows:
\begin{eqnarray}
{\cal O}^{Latt}_1 &=&
[ V_\mu \times P ][ V_\mu \times P] + [ A_\mu \times P ][ A_\mu \times P]
\nonumber \\
{\cal O}^{Latt}_2 &=&
[ V_\mu \times P ][ V_\mu \times P] - [ A_\mu \times P ][ A_\mu \times P]
\nonumber \\
{\cal O}^{Latt}_3 &=& -2 \Big(
[ S \times P ][ S \times P] - [ P \times P ][ P \times P] \Big)
\nonumber \\
{\cal O}^{Latt}_4 &=& 
[ S \times P ][ S \times P] + [ P \times P ][ P \times P]
\nonumber \\
{\cal O}^{Latt}_5 &=& - \frac{1}{2} \Big(
[ S \times P ][ S \times P] + [ P \times P ][ P \times P]
- \sum_{\mu < \nu} [ T_{\mu\nu} \times P ][ T_{\mu\nu} \times P ]
\Big)
\label{eq:latt-basis}
\end{eqnarray}
Here, note that we choose the same flavor structure as the
pseudo-Goldstone mode of the conserved $U(1)_A$ symmetry.

The $SU(4)$ flavor symmetry of the staggered fermion action becomes
exactly conserved only in the limit of $a=0$.
On the finite lattice of $a \ne 0$, only the discrete $SW_{4,diag}$
symmetry and $U(1)_V \otimes U(1)_A$ symmetry are conserved exactly,
which rotates the spin and flavor simultaneously \cite{tool-kit}.
In calculating the perturbative corrections to the operators of
Eq.~(\ref{eq:latt-basis}), the operators with different flavors mix
with one another, since the $SU(4)$ flavor symmetry is not conserved
at $a \ne 0$.
Therefore, we can express a general form of the renormalized
staggered four-fermion operators on the lattice as follows:
\begin{eqnarray}
\vec{\cal O}^{Latt}_i &\equiv& \left( \begin{array}{c} 
			{\cal O}^{Latt}_{i,I} \\ {\cal O}^{Latt}_{i,II}
			\end{array} \right)
\\
\langle \vec{\cal O}^{Latt(1)}_i \rangle &=& 
\langle \vec{\cal O}^{Latt(0)}_i \rangle + \frac{g^2}{(4\pi)^2} \Big[
	- \hat{\Gamma}_{ij} \log( a \kappa ) + 
	\hat{C}^{Latt}_{ij} + \hat{R}_{ij} \Big]
	\langle \vec{\cal O}^{Latt(0)}_j \rangle
\label{eq:ff-latt-Z}
\end{eqnarray}
where $\vec{\cal O}^{Latt(0)}_j$ may include operators with spin or
flavor structure not included in the practical basis of
Eq.~(\ref{eq:latt-basis}).
The anomalous dimension matrices $\hat{\Gamma}_{ij}$ are
scheme-independent and the same as in Eq.~(\ref{eq:ff-anomalous}).
The $\hat{R}_{ij}$ contains the external momentum dependence and will
cancel out when we perform the matching between the continuum and
lattice operators.
The scheme-dependent finite constant matrix terms $\hat{C}^{Latt}_{ij}$
are given in Tables \ref{tab:ff-op-1-I-1}--\ref{tab:ff-op-T-S-P-II-2}.
The analytic formula to calculate these finite constant terms are
provided in Ref.~\cite{JLQCD1}.
Regarding Eq.~(C2) of Ref.~\cite{JLQCD1}, we have obtained a slightly
different analytical result for the corresponding diagram as
follows:
\begin{eqnarray}
G^{3(e)} &=& \frac{g^2}{(4\pi)^2} C_F \delta_{ab'} \delta_{a'b} \Biggl[
\nonumber \\
	&-& 2 Z_{0000} 
	(\overline{\overline{\gamma_{S} \otimes \xi_{F}}})_{CD}
	(\overline{\overline{\gamma_{S'} \otimes \xi_{F'}}})_{C'D'}
\nonumber \\
	&+& \frac{1}{4} Z_{0000} \sum_\mu 
	\Bigl[ (-1)^{(S+F)_\mu} + (-1)^{(S'+F')_\mu} \Bigr]
	(\overline{\overline{\gamma_{\mu 5 S} \otimes \xi_{\mu 5 F}}})_{CD}
	(\overline{\overline{\gamma_{\mu 5 S'} \otimes \xi_{\mu 5 F'}}})_{C'D'}
\nonumber \\
	&+& (1-\alpha) \frac{1}{2} Z_{0000}	
	(\overline{\overline{\gamma_{S} \otimes \xi_{F}}})_{CD}
	(\overline{\overline{\gamma_{S'} \otimes \xi_{F'}}})_{C'D'}
\nonumber \\
	&-& (1-\alpha) \frac{1}{16} Z_{0000}  \sum_\mu
	\Bigl[ (-1)^{(S+F)_\mu} + (-1)^{(S'+F')_\mu} \Bigr]
	(\overline{\overline{\gamma_{\mu 5 S} \otimes \xi_{\mu 5 F}}})_{CD}
	(\overline{\overline{\gamma_{\mu 5 S'} \otimes \xi_{\mu 5 F'}}})_{C'D'}
\nonumber \\
	&+& (1-\alpha) \frac{1}{2} \sum_{\mu \ne \nu, M}
	T^{\mu\nu}_M \biggl\{
\nonumber \\
	& & \qquad
	- \Bigl[ (-1)^{\tilde{M} \cdot (S+F)} + 
	         (-1)^{\tilde{M} \cdot (S'+F')} \Bigr]
	(\overline{\overline{\gamma_{M S} \otimes \xi_{M F}}})_{CD}
	(\overline{\overline{\gamma_{M S'} \otimes \xi_{M F'}}})_{C'D'}
\nonumber \\
	& & \qquad
	+2 \Bigl[ (-1)^{(S+F)_\mu + \tilde{M} \cdot (S+F)} + 
	          (-1)^{(S'+F')_\mu + \tilde{M} \cdot (S'+F')} \Bigr]
\nonumber \\
	& & \qquad \qquad \qquad \times	
(\overline{\overline{\gamma_{\mu 5 M S} \otimes \xi_{\mu 5 M F}}})_{CD}
(\overline{\overline{\gamma_{\mu 5 M S'} \otimes \xi_{\mu 5 M F'}}})_{C'D'}
\nonumber \\
	& & \hspace*{10mm}
	- \Bigl[ (-1)^{(S+F)_\mu + (S+F)_\nu + \tilde{M} \cdot (S+F)} + 
	         (-1)^{(S'+F')_\mu + (S'+F')_\nu + \tilde{M} \cdot (S'+F')} 
	  \Bigr]
\nonumber \\
	& & \qquad \qquad \qquad  \times	
(\overline{\overline{\gamma_{\mu \nu M S} \otimes \xi_{\mu \nu M F}}})_{CD}
(\overline{\overline{\gamma_{\mu \nu M S'} \otimes \xi_{\mu \nu M F'}}})_{C'D'}
	\biggr\}\Biggr]
\label{eq:ff-(e)}
\end{eqnarray}
The difference is localized only to the gauge-dependent terms
proportional to $(1-\alpha)$, the coefficient of the covariant gauge
fixing.
Using Eq.~(\ref{eq:ff-(e)}), we confirmed the gauge invariance by
checking that the summation of all the gauge-dependent terms ({\it
i.e.}  $\propto (1-\alpha)$) of all the one-loop Feynman diagrams
vanishes for each spin and flavor structure of the gauge invariant 
four-fermion operators.

We have checked the results of finite constant terms
$\hat{C}^{Latt}_{ij}$ in three independent ways. 
%
%
First, we checked the conserved $U(1)_A$ symmetry. For example, the
following operators are connected under the $U(1)_A$ transformation
\cite{wlee0}:
\begin{eqnarray}
[ V_\mu \times P ][ V_\mu \times P] 
& \buildrel  U(1)_A  \over  \longrightarrow &
[ V_\mu \times P ][ A_\mu \times S]\,,  \quad 
[ A_\mu \times S ][ V_\mu \times P]\,,  \quad 
[ A_\mu \times S ][ A_\mu \times S]
\nonumber \\ {}
[ A_\mu \times P ][ A_\mu \times P]  
& \buildrel  U(1)_A  \over  \longrightarrow &
[ A_\mu \times P ][ V_\mu \times S] \,, \quad
[ V_\mu \times S ][ A_\mu \times P] \,, \quad
[ V_\mu \times S ][ V_\mu \times S]
\nonumber \\ {}
[ P \times P ][ P \times P ] 
& \buildrel  U(1)_A  \over  \longrightarrow &
[ P \times P ][ S \times S ] \,, \quad
[ S \times S ][ P \times P ] \,, \quad
[ S \times S ][ S \times S ]
\nonumber \\ {}
[ S \times P ][ S \times P ] 
& \buildrel  U(1)_A  \over  \longrightarrow &
[ S \times P ][ P \times S ] \,, \quad
[ P \times S ][ S \times P ] \,, \quad
[ P \times S ][ P \times S ]
\label{eq:U(1)A-identity}
\end{eqnarray}
Hence, we confirmed that all the finite renormalization constants of
the operators in the right-hand side of Eq.~(\ref{eq:U(1)A-identity})
can be reproduced from those in the left-hand side, using a simple
$U(1)_A$ transformation.
Second, we checked the gauge invariance. In other words, we confirmed
that the summation of the gauge dependent terms proportional to $
(1-\alpha) $ vanishes for individual gauge invariant operators in the
practical basis.
Third, we checked the Fierz identities of Eq.~(\ref{eq:fierz-basis}).
Any color one trace operator can be expressed as a linear combination
of color two trace operators, using Fierz transformation, which holds
valid under renormalization on the lattice (but not in the NDR scheme
in the continuum).
We explain the details of these lattice Fierz identities in Appendix
\ref{app:fierz-1} and \ref{app:fierz-2}.

The values of $\hat{C}^{Latt}_{ij}$ presented in Tables
\ref{tab:ff-op-1-I-1}--\ref{tab:ff-op-2-II-2} are in agreement with
those extracted from Table II of Ref. \cite{JLQCD1}.
The values of $\hat{C}^{Latt}_{ij}$ presented in Tables 
\ref{tab:ff-op-3-I-1}--\ref{tab:ff-op-T-S-P-II-2} are new.

We have presented renormalized operators in the continuum and on the
lattice.
We now describe the matching formula.
As in the case of bilinears, the tree level lattice operators are
equal to the tree level continuum operators in the limit of $a=0$.
\begin{eqnarray}
\vec{\cal O}^{Latt(0)} = \vec{\cal O}^{Cont(0)} + O(a) 
\label{eq:ff-tree}
\end{eqnarray}
Using Eqs.~(\ref{eq:ff-cont-Z-2}), (\ref{eq:ff-latt-Z}) and
(\ref{eq:ff-tree}), we can obtain the matching relationship between
continuum operators and lattice operators at one loop level.
\begin{eqnarray}
\vec{\cal O}^{Cont(1)}_i &=& \sum_{j}
\Biggl[ \delta_{ij} 
+ \frac{g^2}{(4\pi)^2} \hat{\Gamma}_{ij} \log( \mu a )
+ \frac{g^2}{(4\pi)^2} \Bigl( \hat{C}^{Cont}_{ij} 
			- \hat{C}^{Latt}_{ij} \Bigr)
\Biggr] \vec{\cal O}^{Latt(1)}_j 
\label{eq:ff-match}
\end{eqnarray}
This is the one-loop matching relation between the lattice and the
continuum four-fermion operators.

We now turn to the tadpole improvement of the staggered four-fermion
operators.
First, let us consider a general color two trace operator:
\begin{eqnarray}
{\cal O}_{II} = \frac{1}{N_f^2} \
[\bar\chi (y_A) 
(\overline{\gamma_S \otimes \xi_F})_{AB} {\cal U}_{AB} \chi(y_B) ] \
[\bar\chi (y_C)
(\overline{\gamma_S' \otimes \xi_F'})_{CD} {\cal U}_{CD} \chi(y_D) ]
\end{eqnarray}
Under the rescaling transformation of Eq.~(\ref{eq:rescale}), this
operator transforms as
\begin{eqnarray}
{\cal O}_{II} \longrightarrow  \tilde{\cal O}_{II}
&=& u_0^{1-\Delta_{SF}} u_0^{1-\Delta_{S'F'}}
{\cal O}_{II}
\end{eqnarray}

In case of color one trace operators, the distance between quark and
anti-quark fields of the bilinears does not define the length of the
gauge link. Therefore it is convenient to express the color one trace
operator as a linear combination of color two trace operators, using
Fierz transformations.
\begin{eqnarray}
{\cal O}^{i}_{I} = \sum_{j} H_{ij} {\cal O}^j_{II}
\end{eqnarray}
where $ H_{ij}$ is the Fierz transformation matrix.
Under the rescaling transformation of Eq.~(\ref{eq:rescale}), 
\begin{eqnarray}
{\cal O}^{i}_{I} \longrightarrow
\tilde{\cal O}^{i}_{I} &=&
\sum_{j} H_{ij} \tilde{\cal O}^j_{II}
\nonumber \\
&=& \sum_{j} H_{ij} 
u_0^{1-\Delta^{j}_{SF}} u_0^{1-\Delta^{j}_{S'F'}} 
{\cal O}^j_{II}
\end{eqnarray}
where $\Delta^{j}_{SF}$ and $\Delta^{j}_{S'F'}$ correspond to
distances between quark and anti-quark fields of the two bilinears in
${\cal O}^j_{II}$ respectively.
Using perturbation, we can rewrite $\tilde{\cal O}^{i}_{I}$ as 
\begin{eqnarray}
\tilde{\cal O}^{i}_{I} &=&
\sum_{j} H_{ij} {\cal O}^j_{II} - 
\frac{g^2}{(4\pi)^2} \frac{(4\pi)^2}{12} \sum_{j} H_{ij} 
\Bigl[ 2 - \Delta^{j}_{SF} - \Delta^{j}_{S'F'} \Bigr]
{\cal O}^j_{II}
\nonumber \\
&=& {\cal O}^{i}_{I} -
\frac{g^2}{(4\pi)^2} \frac{(4\pi)^2}{12} \sum_{j,k} H_{ij} H_{jk} 
 \Bigl[ 2 - \Delta^{j}_{SF} - \Delta^{j}_{S'F'} \Bigr]
{\cal O}^k_{I}
\nonumber \\
&=& {\cal O}^{i}_{I} +
\frac{g^2}{(4\pi)^2} \sum_{k} M_{ik} {\cal O}^k_{I}
\end{eqnarray}
where
\begin{eqnarray}
M_{ik} \equiv - \frac{(4\pi)^2}{12} \sum_j H_{ij} H_{jk}
\Bigl[ 2 - \Delta^{j}_{SF} - \Delta^{j}_{S'F'} \Bigr]
\label{eq:ff-TI-M}
\end{eqnarray}
Here, note that the tadpole improvement for color two trace operators
are diagonal ($\propto \delta_{ij}$), while the tadpole improvement for
color one trace operators allows mixing with off-diagonal operators.
%
%
As a summary, let us express the tadpole-improved four-fermion
operators in terms of unimproved operators as follows:
\begin{eqnarray}
\overrightarrow{ \tilde{\cal O} }^{Latt}_i =
 \left( \begin{array}{c}
	\tilde{\cal O}^{Latt}_{i,I} \\ \tilde{\cal O}^{Latt}_{i,II}
	\end{array} \right)
= \vec{\cal O}^{Latt}_i + 
\frac{g^2}{(4\pi)^2} \sum_j \hat{T}_{ij} \cdot \vec{\cal
O}^{Latt}_j
\label{eq:ff-TI-1}
\end{eqnarray}
where
\begin{eqnarray}
\hat{T}_{ij} &=&  \left( \begin{array}{cc} 
			M_{ij}  &  0  \\ 
			0 &  N_{ij}
			\end{array} \right)
\\
N_{ij} &=& - \delta_{ij} \frac{(4\pi)^2}{12}
\Bigl[ 2 - \Delta^{i}_{SF} - \Delta^{i}_{S'F'} \Bigr]
\label{eq:ff-TI-N}
\end{eqnarray}
The values of $M_{ij}$ and $N_{ij}$ are tabulated in Tables
\ref{tab:ff-op-1-I-1}--\ref{tab:ff-op-T-S-P-II-2}.
Using Eq.~(\ref{eq:ff-TI-1}), one can rewrite the one-loop
renormalization relationship of Eq.~(\ref{eq:ff-latt-Z}) on the
lattice in terms of tadpole-improved operators.
\begin{eqnarray}
\langle \overrightarrow{ \tilde{\cal O} }^{Latt(1)}_i \rangle
	&\equiv&
	\langle \vec{\cal O}^{Latt(0)}_i \rangle + 
	\frac{\tilde{g}^2}{(4\pi)^2} \Big[
	-\hat{\Gamma}_{ij} \log( a \kappa ) + 
	\hat{C}^{Latt}_{ij} + \hat{T}_{ij} +
	\hat{R}_{ij} \Big]
	\langle \vec{\cal O}^{Latt(0)}_j \rangle
\label{eq:ff-latt-Z-TI}
\end{eqnarray}
Here, note that the gauge coupling is also tadpole-improved.
Correspondingly, we need to rewrite the one-loop matching
formula of Eq.~(\ref{eq:ff-match}) in terms of tadpole-improved
operators as follows:
\begin{eqnarray}
\vec{\cal O}^{Cont(1)}_i &=& \sum_{j}
\Biggl[ \delta_{ij} 
+ \frac{\tilde{g}^2}{(4\pi)^2} \hat{\Gamma}_{ij} \log( \mu a )
+ \frac{\tilde{g}^2}{(4\pi)^2} \Bigl( \hat{C}^{Cont}_{ij} 
			- \hat{C}^{Latt}_{ij} - \hat{T}_{ij} \Bigr)
\Biggr] \overrightarrow{ \tilde{\cal O} }^{Latt(1)}_j 
\label{eq:ff-match-TI}
\end{eqnarray}
This is our final result for the tadpole-improved matching relationship
of four-fermion operators.

\section{Conclusion} \label{sec:conclusion}
We have studied the one-loop perturbative matching formula between the
lattice and continuum bilinear and four-fermion operators that are
used in the numerical calculation of $\epsilon'/\epsilon$.
Our main contribution is that the results presented in this paper,
combined with the existing results for the penguin diagrams, make it
possible to match a lattice calculation of $K\rightarrow\pi\pi$ decays
to the continuum NDR results with all the ${\cal O}(g^2)$ corrections
included.
We have studied all five four-fermion operators in the practical
basis.  
The first three operators (${\cal O}_1$, ${\cal O}_2$, and ${\cal
O}_3$) are relevant to the numerical evaluation of
$\epsilon'/\epsilon$.
These operators mix with a large number of off-diagonal operators.
However, we found that the coefficients of the off-diagonal
mixing are small compared with those of the diagonal mixing.
After tadpole-improvement, the diagonal mixing terms are also under
control except for the $([ P \times P ][ P \times P ])_{II}$ term,
which receives a large correction at one loop.
Various attempts to reduce this large perturbative correction of the
$([ P \times P ][ P \times P ])_{II}$ term by improving the action and
operators are under development \cite{wlee&sharpe}.

\section*{Acknowledgments}
The author would like to thank Stephen Sharpe for his consistent
encouragement and helpful discussion on this perturbative calculation.
Valuable discussions with Tanmoy Bhattacharya, Rajan Gupta, and Greg
Kilcup are gratefully acknowledged.
The author would like to express his heartfelt gratitude to Norman
H. Christ and Robert Mawhinney for their support on the
$\epsilon'/\epsilon$ project using staggered fermions.
The author is grateful to Josh Erlich for multiple proof-readings of
this article.
This work was supported in part by DOE grant KA-04-01010-E161.

\appendix
\section{Fierz Transformation (I)} \label{app:fierz-1}
First, we define the notation for Fierz transformation:
\begin{eqnarray}
& & M_S = I \otimes I, \quad
M_V = \sum_{\mu} \gamma_{\mu} \otimes \gamma_{\mu}, \quad
M_T = \sum_{\mu<\nu} \sigma_{\mu\nu} \otimes \sigma_{\mu\nu},
\nonumber
\\
& &
\label{eq:notation}
M_A = \sum_{\mu} \gamma_{\mu5} \otimes \gamma_{\mu5}, \quad
M_P = \gamma_{5} \otimes \gamma_{5}
\end{eqnarray}
where $\otimes$ represents the direct product and $\sigma_{\mu\nu} =
\frac{1}{2} [\gamma_{\mu},\gamma_{\nu}]$ with signature defined by
$\{\gamma_\mu, \gamma_\nu \} = 2 \delta_{\mu\nu}$.
We also introduce the following compact notation for spin indices:
\begin{eqnarray}
(M_V)_{\alpha\beta;\alpha'\beta'}
	= \sum_{\mu} 	(\gamma_{\mu})_{\alpha\beta} \otimes 
			(\gamma_{\mu})_{\alpha'\beta'}
\end{eqnarray}
The same notation also applies to the flavor structure $ \xi_F \otimes
\xi_F $ by switching $ \gamma $ to $ \xi $.
Now we can express the Fierz transformation of $M_X$ for 
$ X \in \{S,V,T,A,P\} $ as follows:
\begin{eqnarray}
\vec{M} = \left( \begin{array}{c} 
		M_S \\  M_V \\  M_T \\ M_A \\ M_P
	  	\end{array} \right) \, ,
\qquad 
\vec{M}_{\alpha\beta';\alpha'\beta} 
= \hat{F} \vec{M}_{\alpha\beta;\alpha'\beta'} \,,
\qquad
(M_i)_{\alpha\beta';\alpha'\beta}
= \left[ F_{ij} \right] \ (M_j )_{\alpha\beta;\alpha'\beta'}
\label{eq:fierz1}
\end{eqnarray}
where
\begin{eqnarray}
\hat{F} = 
\left[ F_{ij} \right]
 =
\frac{1}{4}
\left[
\begin{array}{ccccc}
  1	&  1	&  -1	&  -1	&  1 \\
  4	&  -2	&  0	&  -2	&  -4 \\
  -6	&  0	&  -2	&  0	&  -6  \\
  -4	&  -2	&  0	&  -2	&  4  \\
  1	&  -1	&  -1	&  1	&  1
\end{array}
\right]
\end{eqnarray}

The spin and flavor matrices of the staggered four-fermion operators
are Fierz-transformed separately using the same relationship of
Eq.~(\ref{eq:fierz1}).
Let us define $\Gamma_X$ as the spin part and $\Xi_X$ as the flavor part. 
For example,
\begin{eqnarray*}
[ \Gamma_V \otimes \Xi_P ] &\equiv& [ V_\mu \times P ] [ V_\mu \times P]
\\ {}
[ \Gamma_A \otimes \Xi_P ] &\equiv& [ A_\mu \times P ] [ A_\mu \times P]
\\ {}
[ \Gamma_S \otimes \Xi_P ] &\equiv& [ S \times P ] [ S \times P]
\end{eqnarray*}
Using this notation, we can express a general form of the Fierz
transformation as:
\begin{eqnarray}
[ \Gamma_i \otimes \Xi_m ]^{f_1 f_4 ; f_3 f_2}_{I}
= - F_{ij} F_{mn} [ \Gamma_j \otimes \Xi_n ]^{f_1 f_2 ; f_3 f_4}_{II}
\nonumber \\ {}
[ \Gamma_i \otimes \Xi_m ]^{f_1 f_4 ; f_3 f_2}_{II}
= - F_{ij} F_{mn} [ \Gamma_j \otimes \Xi_n ]^{f_1 f_2 ; f_3 f_4}_{I}
\label{eq:fierz-stag}
\end{eqnarray}
where $I,II$ represent color one trace form and color two trace form 
respectively.
Note that the negative sign on the right-hand side of
Eq.~(\ref{eq:fierz-stag}) comes from the anti-commuting characteristics
of the fermion fields.
Using these relationships of Eqs.~(\ref{eq:fierz1}) and
(\ref{eq:fierz-stag}), we obtain the Fierz transformation of the
practical basis defined in Eq.~(\ref{eq:latt-basis}).
\begin{eqnarray}
[(\Gamma_V + \Gamma_A) \otimes \Xi_P]^{f_1 f_4 ; f_3 f_2}_I &=&
+\frac{1}{4} [ (\Gamma_V + \Gamma_A) \otimes
( \Xi_S - \Xi_V - \Xi_T + \Xi_A + \Xi_P ) ]^{f_1 f_2 ; f_3 f_4}_{II}
\nonumber \\ {}
[(\Gamma_V - \Gamma_A) \otimes \Xi_P]^{f_1 f_4 ; f_3 f_2}_I &=&
- \frac{2}{4} [ (\Gamma_S - \Gamma_P) \otimes
( \Xi_S - \Xi_V - \Xi_T + \Xi_A + \Xi_P ) ]^{f_1 f_2 ; f_3 f_4}_{II}
\nonumber \\ {}
-2 [(\Gamma_S - \Gamma_P) \otimes \Xi_P]^{f_1 f_4 ; f_3 f_2}_I &=&
+\frac{1}{4} [ (\Gamma_V - \Gamma_A) \otimes
( \Xi_S - \Xi_V - \Xi_T + \Xi_A + \Xi_P ) ]^{f_1 f_2 ; f_3 f_4}_{II}
\nonumber \\ {}
[(\Gamma_S + \Gamma_P) \otimes \Xi_P]^{f_1 f_4 ; f_3 f_2}_I &=&
\nonumber \\ {} & & \hspace*{-10mm}
- \frac{1}{8} [ (\Gamma_S + \Gamma_P - \Gamma_T) \otimes
( \Xi_S - \Xi_V - \Xi_T + \Xi_A + \Xi_P ) ]^{f_1 f_2 ; f_3 f_4}_{II}
\nonumber \\ {}
- \frac{1}{2} [(\Gamma_S + \Gamma_P - \Gamma_T) 
		\otimes \Xi_P]^{f_1 f_4 ; f_3 f_2}_I &=&
+\frac{1}{4} [ (\Gamma_S + \Gamma_P) \otimes
( \Xi_S - \Xi_V - \Xi_T + \Xi_A + \Xi_P ) ]^{f_1 f_2 ; f_3 f_4}_{II}
\nonumber \\
\label{eq:fierz-basis}
\end{eqnarray}
Here, note that these Fierz identities hold valid under a
transformation of switching color-trace indices ($I \leftrightarrow
II$) between the left-hand and right-hand sides.
Similarly, we can also derive another Fierz identity used to calculate
$B_K$ in the one spin trace formalism \cite{wlee0}:
\begin{eqnarray}
[ (\Gamma_V + \Gamma_A) \otimes ( \Xi_P + \Xi_S ) ]^{f_1 f_4 ; f_3 f_2}_{I}
= \frac{1}{2}[ (\Gamma_V + \Gamma_A) \times 
(\Xi_S - \Xi_T + \Xi_P ) ]^{f_1 f_2 ; f_3 f_4}_{II}
\end{eqnarray}

\section{Fierz Transformation (II)} \label{app:fierz-2}
As shown in Eq.~(\ref{eq:fierz-basis}), color one trace operators can
be expressed as a linear combination of color two trace operators.
These Fierz identities hold under renormalization on the lattice.
The finite integrals for color one trace operators are completely
different from those for color two trace operators.
Therefore, it will provide a non-trivial check of our calculation
if the results in this paper satisfies the Fierz identities of
Eq.~(\ref{eq:fierz-basis}).

Here, we describe how to check the Fierz identities step by step.
First, we calculate the complete set of one-loop perturbative
corrections for both color one trace and color two trace operators in
Eq.~(\ref{eq:fierz-basis}).
Usually, the operators in the right-hand side of
Eq.~(\ref{eq:fierz-basis}) mix with hundreds of operators at one-loop
level.
In order to compare these results with the operators in the left-hand
side, one needs to Fierz-transform the mixing operators in the
right-hand side into the flavor combination ($f_1f_4;f_3f_2$) as in
the left-hand side.
For this purpose of large scale Fierz transformation, there exists
a handy formula \cite{JLQCD1}.
Using the following group orthogonality relationship for the gamma
matrices:
\begin{eqnarray}
\frac{1}{4} \sum_A  
(\gamma_A)_{\alpha\beta} \otimes (\gamma_A)^\dagger_{\rho\lambda}
= \delta_{\alpha\lambda} \delta_{\beta\rho} \, ,
\end{eqnarray}
one can derive a handy tool for the large scale Fierz
transformation:
\begin{eqnarray}
( \overline{ \gamma_{S } \otimes \xi_{F } } )_{AB}
( \overline{ \gamma_{S'} \otimes \xi_{F'} } )_{A'B'}
= \frac{1}{16} \sum_{KL} 
( \overline{ \gamma_{S } \gamma^\dagger_K 
	\otimes \xi^\dagger_L \xi_{F } } )_{AB'}
( \overline{ \gamma_{S'} \gamma_K \otimes \xi_L \xi_{F'} } )_{A'B} \, .
\label{eq:fierz-handy}
\end{eqnarray}
Using this tool, we confirmed that the lattice results in this paper
satisfy the Fierz identities of Eq.~(\ref{eq:fierz-basis}).

%
%

%

%
\clearpage
%
%
%
\begin{table}[tbh]
\begin{center}
\begin{tabular}{ | l || c | r | }
\hline
${\cal O}_{SF}$  &   $\Gamma_S$  &  $C^{Latt}_{SF;SF}$  \\
\hline 
$ 1 \otimes 1 $                              &     8   &  $+55.628(1)$  \\
$ 1 \otimes \xi_\mu $                        &     8   &  $+14.855(1)$  \\
$ 1 \otimes \xi_{\mu\nu} $                   &     8   &  $-10.578(2)$  \\
$ 1 \otimes \xi_{\mu5} $                     &     8   &  $-29.965(3)$  \\
$ 1 \otimes \xi_{5} $                        &     8   &  $-47.804(2)$  \\
\hline								     
$ \gamma_\mu \otimes 1 $                     &     0   &  $ 0.000(0)$   \\
$ \gamma_\mu \otimes \xi_\mu $               &     0   &  $+19.705(1)$  \\
$ \gamma_\mu \otimes \xi_\nu $               &     0   &  $-13.387(2)$  \\
$ \gamma_\mu \otimes \xi_{\mu\nu} $          &     0   &  $+4.527(1)$   \\
$ \gamma_\mu \otimes \xi_{\nu\rho} $         &     0   &  $-29.654(4)$  \\
$ \gamma_\mu \otimes \xi_{\mu5} $            &     0   &  $-45.998(2)$  \\
$ \gamma_\mu \otimes \xi_{\nu5} $            &     0   &  $-13.411(3)$  \\
$ \gamma_\mu \otimes \xi_5 $                 &     0   &  $-30.010(3)$  \\
\hline								     
$ \gamma_{\mu\nu} \otimes 1 $                &  $-8/3$ &  $-14.620(3)$  \\
$ \gamma_{\mu\nu} \otimes \xi_\mu $          &  $-8/3$ &  $-0.425(1)$   \\
$ \gamma_{\mu\nu} \otimes \xi_\rho $         &  $-8/3$ &  $-29.670(4)$  \\
$ \gamma_{\mu\nu} \otimes \xi_{\mu\nu} $     &  $-8/3$ &  $+7.731(1)$   \\
$ \gamma_{\mu\nu} \otimes \xi_{\mu\rho} $    &  $-8/3$ &  $-14.199(2)$  \\
$ \gamma_{\mu\nu} \otimes \xi_{\rho\sigma} $ &  $-8/3$ &  $-45.396(2)$  \\
\hline
\end{tabular}
\end{center}
\caption{Renormalization constants of bilinear operators at one loop.
${\cal O}_{SF}$ represents bilinear operators. 
$\Gamma_S$ is the anomalous
dimension defined in Eq. (\ref{eq:anomalous1}).
$C^{latt}_{SF;SF}$ is the finite constant term defined in Eq.
(\ref{eq:latt-Z}) for the diagonal case: $S'=S$ and $F'=F$.
}
\label{tab:bi-const-1}
\end{table}
%
%
\begin{table}[tbh]
\begin{center}
\begin{tabular}{ | l | l || r | }
\hline
${\cal O}_{SF}$      &   ${\cal O}_{S'F'}$   &   $C^{latt}_{SF;S'F'}$  \\
\hline 
$ \gamma_\mu \otimes \xi_\nu $	             &
	$ \gamma_\mu \otimes \xi_\mu $       &  $-4.055(0)$  \\
$ \gamma_\mu \otimes \xi_{\mu5} $	     &
	$ \gamma_\mu \otimes \xi_{\nu5} $    &  $+0.862(0)$  \\
$ \gamma_\mu \otimes \xi_{\nu\rho} $	     &
	$ \gamma_\mu \otimes \xi_{\mu\nu} $  &  $+1.982(0)$  \\
$ \gamma_\mu \otimes \xi_{\nu\rho} $	     &
	$ \gamma_\mu \otimes \xi_{\mu\rho} $ &  $-1.982(0)$  \\
\hline								     
$ \gamma_{\mu\nu} \otimes \xi_\rho $         &
	$ \gamma_{\mu\nu} \otimes \xi_\mu $  &  $+0.902(0)$  \\
$ \gamma_{\mu\nu} \otimes \xi_\rho $         &
	$ \gamma_{\mu\nu} \otimes \xi_\nu $  &  $+0.902(0)$  \\
\hline
\end{tabular}
\end{center}
\caption{Renormalization constants of bilinear operators at one loop.
${\cal O}_{SF}$ represents bilinear operators. 
${\cal O}_{S'F'}$ corresponds to the mixing bilinear operators
in Eq. (\ref{eq:bilinearl}). 
$C^{latt}_{SF;S'F'}$ is the finite constant term defined in Eq.
(\ref{eq:latt-Z}) for the off-diagonal case: $S' \ne S$ and $F'\ne F$.
}
\label{tab:bi-const-2}
\end{table}

\clearpage

%
%
%
\begin{table}[h]
\begin{center}
\begin{tabular}{ | l | c || c | r | r | }
\hline 
${\cal O}^{Latt(0)}_j $ 	& color trace 	& $\hat{\Gamma}_{ij}$
 	& $\hat{C}^{Latt}_{ij}$ 	& $ M_{ij} $ \\ \hline
\hline								     
$[ S \times V_\mu ][ S \times V_\mu ] $ 	& I  &  0  &   $-$14.395(3) &  $+c_{TI}$  \\ 
$[ S \times V_\mu ][ S \times V_\mu ] $        	& II &  0  &    $-$3.992(1) &  0 \\ 
$[ S \times V_\mu ][ S \times V_\nu ] $ 	& I  &  0  &    $-$1.110(0) &  0 \\ 
$[ S \times A_\mu ][ S \times A_\mu ] $ 	& I  &  0  &    $-$0.313(0) &  0 \\ 
$[ S \times A_\mu ][ S \times A_\mu ] $ 	& II &  0  &    $-$1.219(0) &  0 \\ 
$[ S \times A_\mu ][ S \times A_\nu ] $		& I  &  0  &    $-$0.119(0) &  0 \\ 
\hline							     
$[ V_\mu \times S ][ V_\mu \times S ] $		& I  &  0  &    +0.342(0) &  0 \\ 
$[ V_\mu \times S ][ V_\mu \times S ] $     	& II &  0  &    $-$0.539(0) &  0 \\ 
$ [ V_\mu \times T_{\mu\nu} ]			
	[ V_\mu \times T_{\mu\nu} ] $ 		& I  &  0  &    +1.397(0) &  0 \\ 
$ [ V_\mu \times T_{\mu\nu} ]			
	[ V_\mu \times T_{\mu\nu} ] $      	& II &  0  &    $-$2.170(0) &  0 \\ 
$ [ V_\mu \times T_{\mu\nu} ]			
	[ V_\mu \times T_{\mu\rho} ] $ 		& I & 0 & +0.565(0) &  0 \\  
$ [ V_\mu \times T_{\mu\nu} ]			
	[ V_\mu \times T_{\mu\rho} ] $	 	& II & 0 & $-$0.499(0) &  0 \\ 
$ [ V_\mu \times T_{\mu\nu} ]			
	[ V_\mu \times T_{\nu\rho} ] $ 		& I & 0 & $-$0.051(0) &  0 \\ 
$ [ V_\mu \times T_{\mu\nu} ]			
	[ V_\mu \times T_{\nu\rho} ] $ 		& II & 0 & +0.154(0) &  0 \\ 
$ [ V_\mu \times T_{\nu\rho} ]			
	[ V_\mu \times T_{\mu\nu} ] $ 		& I & 0 & $-$0.051(0) &  0 \\ 
$ [ V_\mu \times T_{\nu\rho} ]			
	[ V_\mu \times T_{\mu\nu} ] $ 		& II & 0 & +0.154(0) &  0 \\ 
$ [ V_\mu \times T_{\nu\rho} ]			
	[ V_\mu \times T_{\nu\rho} ] $ 		& I & 0 & +1.205(0) &  0 \\ 
$ [ V_\mu \times T_{\nu\rho} ]			
	[ V_\mu \times T_{\nu\rho} ] $ 		& II & 0 & $-$1.972(0) &  0 \\ 
$ [ V_\mu \times T_{\nu\rho} ]			
	[ V_\mu \times T_{\nu\eta} ] $ 		& I & 0 & +0.502(0) &  0 \\ 
$ [ V_\mu \times T_{\nu\rho} ]			
	[ V_\mu \times T_{\nu\eta} ] $ 		& II & 0 & $-$0.307(0) &  0 \\ 
$ [ V_\mu \times P ]
	[ V_\mu \times P ] $ 	& I & $+2$ & $-$24.167(2) &  $+2 c_{TI}$ \\  
$ [ V_\mu \times P ]
	[ V_\mu \times P ] $ 	& II & $-6$ & $-$7.028(3) &  0 \\ 
\hline
$ [ T_{\mu\nu} \times V_\mu ]
	[ T_{\mu\nu} \times V_\mu ] $ 		& I & 0 & +18.270(2) &  $-c_{TI}$ \\ 
$ [ T_{\mu\nu} \times V_\mu ]
	[ T_{\mu\nu} \times V_\mu ] $  		& II & 0 & $-$4.544(1) &  0 \\ 
$ [ T_{\mu\nu} \times V_\mu ]
	[ T_{\mu\nu} \times V_\nu ] $ 		& I & 0 & $-$1.952(0) &  0 \\ 
$ [ T_{\mu\nu} \times V_\mu ]
	[ T_{\mu\nu} \times V_\nu ] $ 		& II & 0 & +2.527(0) &  0 \\ 
$ [ T_{\mu\nu} \times V_\mu ]
	[ T_{\mu\nu} \times V_\rho ] $ 		& I & 0 & $-$0.223(0) &  0 \\ 
$ [ T_{\mu\nu} \times V_\mu ]
	[ T_{\mu\nu} \times V_\rho ] $  	& II & 0 & +0.669(0) &  0 \\ 
$ [ T_{\mu\nu} \times V_\rho ]
	[ T_{\mu\nu} \times V_\mu ] $ 		& I & 0 & $-$0.223(0) &  0 \\ 
$ [ T_{\mu\nu} \times V_\rho ]
	[ T_{\mu\nu} \times V_\mu ] $  		& II & 0 & +0.669(0) &  0 \\ 
$ [ T_{\mu\nu} \times V_\rho ]
	[ T_{\mu\nu} \times V_\rho ] $ 		& I & 0 & $-$15.294(2) &  $+c_{TI}$ \\ 
$ [ T_{\mu\nu} \times V_\rho ]
	[ T_{\mu\nu} \times V_\rho ] $ 		& II & 0 & $-$4.387(1) &  0 \\ 
$ [ T_{\mu\nu} \times V_\rho ]
	[ T_{\mu\nu} \times V_\eta ] $ 		& I & 0 & +1.110(0) &  0 \\ 
\hline
\end{tabular}
\end{center}
\caption{Renormalization constants of a color one trace four-fermion
operator at one loop, $({\cal O}^{Latt}_1)_{I}$ defined in Eq. 
(\ref{eq:latt-basis}).
%
%
The $\hat{\Gamma}_{ij}$ matrix represents an anomalous dimension
defined in Eq.~(\ref{eq:ff-latt-Z}) and its values are given in
Eq.~(\ref{eq:ff-anomalous}).
The $\hat{C}^{Latt}_{ij}$ matrix is defined in
Eq.~(\ref{eq:ff-latt-Z}).
The $M_{ij}$ matrix represents the tadpole-improvement
defined in Eq.~(\ref{eq:ff-TI-M}).
$c_{TI} \equiv (4\pi)^2 / 12$.
All the greek indices are summed under the condition of
$\mu \ne \nu \ne \rho \ne \eta$.
}
\label{tab:ff-op-1-I-1}
\end{table}
\begin{table}[h]
\begin{center}
\begin{tabular}{ | l | c || c | r | r | }
\hline
${\cal O}^{Latt(0)}_j $ 	& color trace 	& $\hat{\Gamma}_{ij}$
 	& $\hat{C}^{Latt}_{ij}$ 	& $ M_{ij} $ \\ \hline
\hline
$ [ T_{\mu\nu} \times A_\mu ]
	[ T_{\mu\nu} \times A_\mu ] $ 		& I & 0 & +0.018(0) &  0 \\ 
$ [ T_{\mu\nu} \times A_\mu ]
	[ T_{\mu\nu} \times A_\mu ] $  		& II & 0 & $-$1.061(0) &  0 \\ 
$ [ T_{\mu\nu} \times A_\mu ]
	[ T_{\mu\nu} \times A_\nu ] $ 		& I & 0 & $-$0.055(0) &  0 \\ 
$ [ T_{\mu\nu} \times A_\mu ]
	[ T_{\mu\nu} \times A_\nu ] $  		& II & 0 & $-$0.191(0) &  0 \\ 
$ [ T_{\mu\nu} \times A_\mu ]
	[ T_{\mu\nu} \times A_\rho ] $ 		& I & 0 & +0.032(0) &  0 \\ 
$ [ T_{\mu\nu} \times A_\mu ]
	[ T_{\mu\nu} \times A_\rho ] $  	& II & 0 & $-$0.096(0) &  0 \\ 
$ [ T_{\mu\nu} \times A_\rho ]
	[ T_{\mu\nu} \times A_\mu ] $ 		& I & 0 & +0.032(0) &  0 \\ 
$ [ T_{\mu\nu} \times A_\rho ]
	[ T_{\mu\nu} \times A_\mu ] $  		& II & 0 & $-$0.096(0) &  0 \\ 
$ [ T_{\mu\nu} \times A_\rho ]
	[ T_{\mu\nu} \times A_\rho ] $ 		& I & 0 & +0.742(0) &  0 \\ 
$ [ T_{\mu\nu} \times A_\rho ]
	[ T_{\mu\nu} \times A_\rho ] $ 		& II & 0 & $-$1.219(0) &  0 \\ 
$ [ T_{\mu\nu} \times A_\rho ]
	[ T_{\mu\nu} \times A_\eta ] $ 		& I & 0 & +0.119(0) &  0 \\ 
\hline 
$ [ A_\mu \times S ][ A_\mu \times S ] $ 	& I & 0 & +0.342(0) &  0 \\ 
$ [ A_\mu \times S ][ A_\mu \times S ] $ 	& II & 0 & $-$0.539(0) &  0 \\ 
$ [ A_\mu \times T_{\mu\nu} ]
	[ A_\mu \times T_{\mu\nu} ] $ 		& I & 0 & +1.397(0) &  0 \\ 
$ [ A_\mu \times T_{\mu\nu} ]
	[ A_\mu \times T_{\mu\nu} ] $ 		& II & 0 & $-$2.170(0) &  0 \\ 
$ [ A_\mu \times T_{\mu\nu} ]
	[ A_\mu \times T_{\mu\rho} ] $ 		& I & 0 & +0.399(0) &  0 \\ 
$ [ A_\mu \times T_{\mu\nu} ]
	[ A_\mu \times T_{\nu\rho} ] $ 		& I & 0 & $-$0.083(0) &  0 \\ 
$ [ A_\mu \times T_{\mu\nu} ]
	[ A_\mu \times T_{\nu\rho} ] $ 		& II & 0 & +0.249(0) &  0 \\ 
$ [ A_\mu \times T_{\nu\rho} ]
	[ A_\mu \times T_{\mu\nu} ] $ 		& I & 0 & $-$0.083(0) &  0 \\ 
$ [ A_\mu \times T_{\nu\rho} ]
	[ A_\mu \times T_{\mu\nu} ] $ 		& II & 0 & +0.249(0) &  0 \\ 
$ [ A_\mu \times T_{\nu\rho} ]
	[ A_\mu \times T_{\nu\rho} ] $ 		& I & 0 & +1.336(0) &  0 \\ 
$ [ A_\mu \times T_{\nu\rho} ]
	[ A_\mu \times T_{\nu\rho} ] $  	& II & 0 & $-$2.367(0) &  0 \\ 
$ [ A_\mu \times T_{\nu\rho} ]
	[ A_\mu \times T_{\nu\eta} ] $ 	& I & 0 & +0.399(0) &  0 \\ 
$ [ A_\mu \times P ]
	[ A_\mu \times P ] $ 	& I & $+2$ & $-$24.768(2) & $+2c_{TI}$ \\ 
$ [ A_\mu \times P ]
	[ A_\mu \times P ] $ 	& II & $-6$ & $-$5.225(1) &  0 \\ 
\hline
$ [ P \times V_\mu ][ P \times V_\mu ] $ 	& I & 0 & +17.108(2) &  $-c_{TI}$ \\ 
$ [ P \times V_\mu ][ P \times V_\mu ] $ 	& II & 0 & $-$4.149(1) &  0 \\  
$ [ P \times V_\mu ][ P \times V_\nu ] $ 	& I & 0 & +0.805(0) &  0 \\ 
$ [ P \times V_\mu ][ P \times V_\nu ] $ 	& II & 0 & +0.915(0) &  0 \\ 
$ [ P \times A_\mu ][ P \times A_\mu ] $ 	& I & 0 & +1.073(0) &  0 \\ 
$ [ P \times A_\mu ][ P \times A_\mu ] $ 	& II & 0 & $-$1.061(0) &  0 \\ 
$ [ P \times A_\mu ][ P \times A_\nu ] $ 	& I & 0 & +0.183(0) &  0 \\ 
$ [ P \times A_\mu ][ P \times A_\nu ] $ 	& II & 0 & $-$0.191(0) &  0 \\ 
\hline
\end{tabular}
\end{center}
\caption{Renormalization constants of a color one trace four-fermion
operator at one loop, $({\cal O}^{Latt}_1)_{I}$ defined in Eq. 
(\ref{eq:latt-basis}).
%
%
The $\hat{\Gamma}_{ij}$ matrix represents an anomalous dimension
defined in Eq.~(\ref{eq:ff-latt-Z}) and its values are given in
Eq.~(\ref{eq:ff-anomalous}).
The $\hat{C}^{Latt}_{ij}$ matrix is defined in
Eq.~(\ref{eq:ff-latt-Z}).
The $M_{ij}$ matrix represents the tadpole-improvement
defined in Eq.~(\ref{eq:ff-TI-M}).
$c_{TI} \equiv (4\pi)^2 / 12$.
All the greek indices are summed under the condition of
$\mu \ne \nu \ne \rho \ne \eta$.
}
\label{tab:ff-op-1-I-2}
\end{table}

\clearpage

%
%
%
\begin{table}[h]
\begin{center}
\begin{tabular}{ | l | c || c | r | r | }
\hline 
${\cal O}^{Latt(0)}_j $ 	& color trace 	& $\hat{\Gamma}_{ij}$
 	& $\hat{C}^{Latt}_{ij}$ 	& $ N_{ij} $ \\ \hline
\hline								     
$ [ S \times V_\mu ][ S \times V_\mu ] $ & I & 0 & $-5.581$(1) &  0 \\ 
$ [ S \times V_\mu ][ S \times V_\mu ] $ & II & 0 & $+1.860$(0) &  0 \\ 
$ [ S \times V_\mu ][ S \times V_\nu ] $ & I & 0 & $+1.114$(0) &  0 \\ 
$ [ S \times V_\mu ][ S \times V_\nu ] $ & II & 0 & $-0.371$(0) &  0 \\ 
$ [ S \times A_\mu ][ S \times A_\mu ] $ & I & 0 & $-1.311$(0) &  0 \\ 
$ [ S \times A_\mu ][ S \times A_\mu ] $ & II & 0 & $+0.437$(0) &  0 \\ 
$ [ S \times A_\mu ][ S \times A_\nu ] $ & I & 0 & $-0.236$(0) &  0 \\ 
$ [ S \times A_\mu ][ S \times A_\nu ] $ & II & 0 & $+0.079$(0) &  0 \\ 
\hline
$ [ V_\mu \times S ][ V_\mu \times S ] $ & I & 0 & $-1.107$(0) &  0 \\ 
$ [ V_\mu \times S ][ V_\mu \times S ] $ & II & 0 & $+0.369$(0) &  0 \\ 
$ [ V_\mu \times T_{\mu\nu} ]
	[ V_\mu \times T_{\mu\nu} ] $ & I & 0 & $-1.106$(0) &  0 \\ 
$ [ V_\mu \times T_{\mu\nu} ]
	[ V_\mu \times T_{\mu\nu} ] $ & II & 0 & $+0.369$(0) &  0 \\ 
$ [ V_\mu \times T_{\mu\nu} ]
	[ V_\mu \times T_{\mu\rho} ] $ & I & 0 & $+0.150$(0) &  0 \\ 
$ [ V_\mu \times T_{\mu\nu} ]
	[ V_\mu \times T_{\mu\rho} ] $ & II & 0 & $-0.050$(0) &  0 \\ 
$ [ V_\mu \times T_{\mu\nu} ]
	[ V_\mu \times T_{\nu\rho} ] $ & I & 0 & $+0.403$(0) &  0 \\ 
$ [ V_\mu \times T_{\mu\nu} ]
	[ V_\mu \times T_{\nu\rho} ] $ & II & 0 & $-0.134$(0) &  0 \\ 
$ [ V_\mu \times T_{\nu\rho} ]
	[ V_\mu \times T_{\mu\nu} ] $ & I & 0 & $+0.403$(0) &  0 \\ 
$ [ V_\mu \times T_{\nu\rho} ] 
	[ V_\mu \times T_{\mu\nu} ] $ & II & 0 & $-0.134$(0) &  0 \\ 
$ [ V_\mu \times T_{\nu\rho} ]
	[ V_\mu \times T_{\nu\rho} ] $ & I & 0 & $-2.776$(0) &  0 \\ 
$ [ V_\mu \times T_{\nu\rho} ]
	[ V_\mu \times T_{\nu\rho} ] $ & II & 0 & $+0.925$(0) &  0 \\ 
$ [ V_\mu \times T_{\nu\rho} ]
	[ V_\mu \times T_{\nu\eta} ] $ & I & 0 & $+0.150$(0) &  0 \\ 
$ [ V_\mu \times T_{\nu\rho} ]
	[ V_\mu \times T_{\nu\eta} ] $ & II & 0 & $-0.050$(0) &  0 \\ 
$ [ V_\mu \times P ]
	[ V_\mu \times P ] $ & I & $-6$ & $-4.500$(0) &  0 \\ 
$ [ V_\mu \times P ]
	[ V_\mu \times P ] $ & II & $+2$ & $-58.523$(7) & $+4 c_{TI}$  \\ 
\hline
$ [ T_{\mu\nu} \times V_\mu ]
	[ T_{\mu\nu} \times V_\mu ] $ & I & 0 & $-2.767$(0) &  0 \\ 
$ [ T_{\mu\nu} \times V_\mu ]
	[ T_{\mu\nu} \times V_\mu ] $ & II & 0 & $+0.922$(0) &  0 \\ 
$ [ T_{\mu\nu} \times V_\mu ]
	[ T_{\mu\nu} \times V_\nu ] $ & I & 0 & $-0.416$(0) &  0 \\ 
$ [ T_{\mu\nu} \times V_\mu ]
	[ T_{\mu\nu} \times V_\nu ] $ & II & 0 & $+0.139$(0) &  0 \\ 
$ [ T_{\mu\nu} \times V_\mu ]
	[ T_{\mu\nu} \times V_\rho ] $ & I & 0 & $+0.765$(0) &  0 \\ 
$ [ T_{\mu\nu} \times V_\mu ]
	[ T_{\mu\nu} \times V_\rho ] $ & II & 0 & $-0.255$(0) &  0 \\ 
$ [ T_{\mu\nu} \times V_\rho ]
	[ T_{\mu\nu} \times V_\mu ] $ & I & 0 & $+0.765$(0) &  0 \\ 
$ [ T_{\mu\nu} \times V_\rho ]
	[ T_{\mu\nu} \times V_\mu ] $ & II & 0 & $-0.255$(0) &  0 \\ 
$ [ T_{\mu\nu} \times V_\rho ]
	[ T_{\mu\nu} \times V_\rho ] $ & I & 0 & $-5.968$(1) &  0 \\ 
$ [ T_{\mu\nu} \times V_\rho ]
	[ T_{\mu\nu} \times V_\rho ] $ & II & 0 & $+1.989$(0) &  0 \\ 
$ [ T_{\mu\nu} \times V_\rho ]
	[ T_{\mu\nu} \times V_\eta ] $ & I & 0 & $+1.946$(0) &  0 \\ 
$ [ T_{\mu\nu} \times V_\rho ]
	[ T_{\mu\nu} \times V_\eta ] $ & II & 0 & $-0.649$(0) &  0 \\ 
\hline
\end{tabular}
\end{center}
\caption{Renormalization constants of a color two trace four-fermion
operator at one loop, $({\cal O}^{Latt}_1)_{II}$ defined in Eq. 
(\ref{eq:latt-basis}).
%
%
The $\hat{\Gamma}_{ij}$ matrix represents an anomalous dimension
defined in Eq.~(\ref{eq:ff-latt-Z}) and its values are given in
Eq.~(\ref{eq:ff-anomalous}).
The $\hat{C}^{Latt}_{ij}$ matrix is defined in
Eq.~(\ref{eq:ff-latt-Z}).
The $N_{ij}$ matrix represents the tadpole-improvement
defined in Eq.~(\ref{eq:ff-TI-N}).
$c_{TI} \equiv (4\pi)^2 / 12$.
All the greek indices are summed under the condition of
$\mu \ne \nu \ne \rho \ne \eta$.
}
\label{tab:ff-op-1-II-1}
\end{table}
\begin{table}[h]
\begin{center}
\begin{tabular}{ | l | c || c | r | r | }
\hline
${\cal O}^{Latt(0)}_j $ 	& color trace 	& $\hat{\Gamma}_{ij}$
 	& $\hat{C}^{Latt}_{ij}$ 	& $ N_{ij} $ \\ \hline
\hline 
$ [ T_{\mu\nu} \times A_\mu ]
	[ T_{\mu\nu} \times A_\mu ] $ & I & 0 & $-0.772$(0) &  0 \\ 
$ [ T_{\mu\nu} \times A_\mu ]
	[ T_{\mu\nu} \times A_\mu ] $ & II & 0 & $+0.257$(0) &  0 \\ 
$ [ T_{\mu\nu} \times A_\mu ]
	[ T_{\mu\nu} \times A_\nu ] $ & I & 0 & $-0.045$(0) &  0 \\ 
$ [ T_{\mu\nu} \times A_\mu ]
	[ T_{\mu\nu} \times A_\nu ] $ & II & 0 & $+0.015$(0) &  0 \\ 
$ [ T_{\mu\nu} \times A_\rho ]
	[ T_{\mu\nu} \times A_\rho ] $ & I & 0 & $-1.705$(0) &  0 \\ 
$ [ T_{\mu\nu} \times A_\rho ]
	[ T_{\mu\nu} \times A_\rho ] $ & II & 0 & $+0.568$(0) &  0 \\ 
$ [ T_{\mu\nu} \times A_\rho ]
	[ T_{\mu\nu} \times A_\eta ] $ & I & 0 & $-0.529$(0) &  0 \\ 
$ [ T_{\mu\nu} \times A_\rho ]
	[ T_{\mu\nu} \times A_\eta ] $ & II & 0 & $+0.176$(0) &  0 \\ 
\hline
$ [ A_\mu \times S ][ A_\mu \times S ] $ & I & 0 & $-0.478$(0) &  0 \\ 
$ [ A_\mu \times S ][ A_\mu \times S ] $ & II & 0 & $+0.159$(0) &  0 \\ 
$ [ A_\mu \times T_{\mu\nu} ]
	[ A_\mu \times T_{\mu\nu} ] $ & I & 0 & $-1.106$(0) &  0 \\ 
$ [ A_\mu \times T_{\mu\nu} ]
	[ A_\mu \times T_{\mu\nu} ] $ & II & 0 & $+0.369$(0) &  0 \\ 
$ [ A_\mu \times T_{\mu\nu} ]
	[ A_\mu \times T_{\mu\rho} ] $ & I & 0 & $+0.150$(0) &  0 \\ 
$ [ A_\mu \times T_{\mu\nu} ]
	[ A_\mu \times T_{\mu\rho} ] $ & II & 0 & $-0.050$(0) &  0 \\ 
$ [ A_\mu \times T_{\nu\rho} ]
	[ A_\mu \times T_{\nu\rho} ] $ & I & 0 & $-2.776$(0) &  0 \\ 
$ [ A_\mu \times T_{\nu\rho} ]
	[ A_\mu \times T_{\nu\rho} ] $ & II & 0 & $+0.925$(0) &  0 \\ 
$ [ A_\mu \times T_{\nu\rho} ]
	[ A_\mu \times T_{\nu\eta} ] $ & I & 0 & $-0.656$(0) &  0 \\ 
$ [ A_\mu \times T_{\nu\rho} ]
	[ A_\mu \times T_{\nu\eta} ] $ & II & 0 & $+0.219$(0) &  0 \\ 
$ [ A_\mu \times P ][ A_\mu \times P ] $ & I & $-6$ & $-4.500$(0) &  0 \\ 
$ [ A_\mu \times P ][ A_\mu \times P ] $ & II & $+2$ & $+1.501$(2) &  0 \\ 
\hline
$ [ P \times V_\mu ][ P \times V_\mu ] $ & I & 0 & $-3.153$(1) &  0 \\ 
$ [ P \times V_\mu ][ P \times V_\mu ] $ & II & 0 & $+1.051$(0) &  0 \\ 
$ [ P \times V_\mu ][ P \times V_\nu ] $ & I & 0 & $+0.416$(0) &  0 \\ 
$ [ P \times V_\mu ][ P \times V_\nu ] $ & II & 0 & $-0.139$(0) &  0 \\ 
$ [ P \times A_\mu ][ P \times A_\mu ] $ & I & 0 & $-0.377$(0) &  0 \\ 
$ [ P \times A_\mu ][ P \times A_\mu ] $ & II & 0 & $+0.126$(0) &  0 \\ 
$ [ P \times A_\mu ][ P \times A_\nu ] $ & I & 0 & $+0.045$(0) &  0 \\ 
$ [ P \times A_\mu ][ P \times A_\nu ] $ & II & 0 & $-0.015$(0) &  0 \\ 
\hline
\end{tabular}
\end{center}
\caption{Renormalization constants of a color two trace four-fermion
operator at one loop, $({\cal O}^{Latt}_1)_{II}$ defined in Eq. 
(\ref{eq:latt-basis}).
%
%
The $\hat{\Gamma}_{ij}$ matrix represents an anomalous dimension
defined in Eq.~(\ref{eq:ff-latt-Z}) and its values are given in
Eq.~(\ref{eq:ff-anomalous}).
The $\hat{C}^{Latt}_{ij}$ matrix is defined in
Eq.~(\ref{eq:ff-latt-Z}).
The $N_{ij}$ matrix represents the tadpole-improvement
defined in Eq.~(\ref{eq:ff-TI-N}).
$c_{TI} \equiv (4\pi)^2 / 12$.
All the greek indices are summed under the condition of
$\mu \ne \nu \ne \rho \ne \eta$.
}
\label{tab:ff-op-1-II-2}
\end{table}

\clearpage

%
%
%
\begin{table}[h]
\begin{center}
\begin{tabular}{ | l | c || c | r | r | }
\hline 
${\cal O}^{Latt(0)}_j $ 	& color trace 	& $\hat{\Gamma}_{ij}$
 	& $\hat{C}^{Latt}_{ij}$ 	& $ M_{ij} $ \\ \hline
\hline								     

$ [ S \times V_\mu ][ S \times V_\mu ] $ & I & 0 & $+23.853$(3) & $-c_{TI}$  \\ 
$ [ S \times V_\mu ][ S \times V_\mu ] $ & II & 0 & $+0.839$(1) & 0  \\ 
$ [ S \times V_\mu ][ S \times V_\nu ] $ & I & 0 & $-0.139$(0) &  0 \\ 
$ [ S \times V_\mu ][ S \times V_\nu ] $ & II & 0 & $+0.416$(0) & 0  \\ 
$ [ S \times A_\mu ][ S \times A_\mu ] $ & I & 0 & $-1.443$(0) &  0 \\ 
$ [ S \times A_\mu ][ S \times A_\mu ] $ & II & 0 & $-0.842$(0) & 0  \\ 
$ [ S \times A_\mu ][ S \times A_\nu ] $ & I & 0 & $+0.015$(0) &  0 \\ 
$ [ S \times A_\mu ][ S \times A_\nu ] $ & II & 0 & $-0.045$(0) & 0 \\ 
\hline
$ [ V_\mu \times S ][ V_\mu \times S ] $ & I & 0 & $+0.515$(0) &  0 \\ 
$ [ V_\mu \times S ][ V_\mu \times S ] $ & II & 0 & $+0.254$(0) & 0  \\ 
$ [ V_\mu \times T_{\mu\nu} ]
	[ V_\mu \times T_{\mu\nu} ] $ & I & 0 & $-3.977$(0) &  0 \\ 
$ [ V_\mu \times T_{\mu\nu} ]
	[ V_\mu \times T_{\mu\nu} ] $ & II & 0 & $+1.064$(0) &  0 \\ 
$ [ V_\mu \times T_{\mu\nu} ]
	[ V_\mu \times T_{\mu\rho} ] $ & I & 0 & $-0.216$(0) &  0 \\ 
$ [ V_\mu \times T_{\mu\nu} ]
	[ V_\mu \times T_{\mu\rho} ] $ & II & 0 & $+0.648$(0) & 0 \\ 
$ [ V_\mu \times T_{\mu\nu} ]
	[ V_\mu \times T_{\nu\rho} ] $ & I & 0 & $-0.051$(0) & 0 \\ 
$ [ V_\mu \times T_{\mu\nu} ]
	[ V_\mu \times T_{\nu\rho} ] $ & II & 0 & $+0.154$(0) & 0 \\ 
$ [ V_\mu \times T_{\nu\rho} ]
	[ V_\mu \times T_{\mu\nu} ] $ & I & 0 & $-0.051$(0) &  0 \\ 
$ [ V_\mu \times T_{\nu\rho} ]
	[ V_\mu \times T_{\mu\nu} ] $ & II & 0 & $+0.154$(0) & 0 \\ 
$ [ V_\mu \times T_{\nu\rho} ]
	[ V_\mu \times T_{\nu\rho} ] $ & I & 0 & $-3.486$(0) & 0 \\ 
$ [ V_\mu \times T_{\nu\rho} ]
	[ V_\mu \times T_{\nu\rho} ] $ & II & 0 & $-0.409$(0) & 0 \\ 
$ [ V_\mu \times T_{\nu\rho} ]
	[ V_\mu \times T_{\nu\eta} ] $ & I & 0 & $+0.053$(0) & 0 \\ 
$ [ V_\mu \times T_{\nu\rho} ]
	[ V_\mu \times T_{\nu\eta} ] $ & II & 0 & $-0.158$(0) & 0 \\ 
$ [ V_\mu \times P ][ V_\mu \times P ] $ & I & $+16$ & $-13.668$(2) & $+2c_{TI}$ \\ 
$ [ V_\mu \times P ][ V_\mu \times P ] $ & II & 0 & $-2.529$(3) & 0 \\ 
\hline
$ [ T_{\mu\nu} \times V_\mu ]
	[ T_{\mu\nu} \times V_\mu ] $ & I & 0 & $-24.725$(2) & $+c_{TI}$ \\ 
$ [ T_{\mu\nu} \times V_\mu ]
	[ T_{\mu\nu} \times V_\mu ] $ & II & 0 & $+1.778$(1) & 0 \\ 
$ [ T_{\mu\nu} \times V_\mu ]
	[ T_{\mu\nu} \times V_\nu ] $ & I & 0 & $+0.981$(0) &  0 \\ 
$ [ T_{\mu\nu} \times V_\mu ]
	[ T_{\mu\nu} \times V_\nu ] $ & II & 0 & $-2.943$(0) &  0 \\ 
$ [ T_{\mu\nu} \times V_\mu ]
	[ T_{\mu\nu} \times V_\rho ] $ & I & 0 & $-0.223$(0) &  0 \\ 
$ [ T_{\mu\nu} \times V_\mu ]
	[ T_{\mu\nu} \times V_\rho ] $ & II & 0 & $+0.669$(0) &  0 \\ 
$ [ T_{\mu\nu} \times V_\rho ]
	[ T_{\mu\nu} \times V_\mu ] $ & I & 0 & $-0.223$(0) &  0  \\ 
$ [ T_{\mu\nu} \times V_\rho ]
	[ T_{\mu\nu} \times V_\mu ] $ & II & 0 & $+0.669$(0) &  0 \\ 
$ [ T_{\mu\nu} \times V_\rho ]
	[ T_{\mu\nu} \times V_\rho ] $ & I & 0 & $-23.594$(2) & 
							$+c_{TI}$ \\ 
$ [ T_{\mu\nu} \times V_\rho ]
	[ T_{\mu\nu} \times V_\rho ] $ & II & 0 & $-1.621$(1) &  0 \\ 
$ [ T_{\mu\nu} \times V_\rho ]
	[ T_{\mu\nu} \times V_\eta ] $ & I & 0 & $-0.139$(0) &  0 \\ 
$ [ T_{\mu\nu} \times V_\rho ]
	[ T_{\mu\nu} \times V_\eta ] $ & II & 0 & $+0.416$(0) & 0 \\ 
\hline
\end{tabular}
\end{center}
\caption{Renormalization constants of a color one trace four-fermion
operator at one loop, $({\cal O}^{Latt}_2)_{I}$ defined in Eq. 
(\ref{eq:latt-basis}). 
%
%
The $\hat{\Gamma}_{ij}$ matrix represents an anomalous dimension
defined in Eq.~(\ref{eq:ff-latt-Z}) and its values are given in
Eq.~(\ref{eq:ff-anomalous}).
The $\hat{C}^{Latt}_{ij}$ matrix is defined in
Eq.~(\ref{eq:ff-latt-Z}).
The $M_{ij}$ matrix represents the tadpole-improvement
defined in Eq.~(\ref{eq:ff-TI-M}).
$c_{TI} \equiv (4\pi)^2 / 12$.
All the greek indices are summed under the condition of
$\mu \ne \nu \ne \rho \ne \eta$.
}
\label{tab:ff-op-2-I-1}
\end{table}
\begin{table}[h]
\begin{center}
\begin{tabular}{ | l | c || c | r | r | }
\hline
${\cal O}^{Latt(0)}_j $ 	& color trace 	& $\hat{\Gamma}_{ij}$
 	& $\hat{C}^{Latt}_{ij}$ 	& $ M_{ij} $ \\ \hline
\hline 
$ [ T_{\mu\nu} \times A_\mu ]
	[ T_{\mu\nu} \times A_\mu ] $ & I & 0 & $+1.820$(0) &  0 \\ 
$ [ T_{\mu\nu} \times A_\mu ]
	[ T_{\mu\nu} \times A_\mu ] $ & II & 0 & $-0.289$(0) &  0 \\ 
$ [ T_{\mu\nu} \times A_\mu ]
	[ T_{\mu\nu} \times A_\nu ] $ & I & 0 & $+0.049$(0) &  0 \\ 
$ [ T_{\mu\nu} \times A_\mu ]
	[ T_{\mu\nu} \times A_\nu ] $ & II & 0 & $-0.146$(0) &  0 \\ 
$ [ T_{\mu\nu} \times A_\mu ]
	[ T_{\mu\nu} \times A_\rho ] $ & I & 0 & $+0.032$(0) &  0 \\ 
$ [ T_{\mu\nu} \times A_\mu ]
	[ T_{\mu\nu} \times A_\rho ] $ & II & 0 & $-0.096$(0) & 0  \\ 
$ [ T_{\mu\nu} \times A_\rho ]
	[ T_{\mu\nu} \times A_\mu ] $ & I & 0 & $+0.032$(0) &  0 \\ 
$ [ T_{\mu\nu} \times A_\rho ]
	[ T_{\mu\nu} \times A_\mu ] $ & II & 0 & $-0.096$(0) &  0 \\ 
$ [ T_{\mu\nu} \times A_\rho ]
	[ T_{\mu\nu} \times A_\rho ] $ & I & 0 & $+1.575$(0) &  0 \\ 
$ [ T_{\mu\nu} \times A_\rho ]
	[ T_{\mu\nu} \times A_\rho ] $ & II & 0 & $+0.446$(0) & 0 \\ 
$ [ T_{\mu\nu} \times A_\rho ]
	[ T_{\mu\nu} \times A_\eta ] $ & I & 0 & $+0.015$(0) & 0 \\ 
$ [ T_{\mu\nu} \times A_\rho ]
	[ T_{\mu\nu} \times A_\eta ] $ & II & 0 & $-0.045$(0) & 0 \\ 
\hline
$ [ A_\mu \times S ][ A_\mu \times S ] $ & I & 0 & $-0.515$(0) & 0 \\ 
$ [ A_\mu \times S ][ A_\mu \times S ] $ & II & 0 & $-0.254$(0) & 0 \\ 
$ [ A_\mu \times T_{\mu\nu} ]
	[ A_\mu \times T_{\mu\nu} ] $ & I & 0 & $+3.977$(0) & 0 \\ 
$ [ A_\mu \times T_{\mu\nu} ]
	[ A_\mu \times T_{\mu\nu} ] $ & II & 0 & $-1.064$(0) & 0 \\ 
$ [ A_\mu \times T_{\mu\nu} ]
	[ A_\mu \times T_{\mu\rho} ] $ & I & 0 & $+0.050$(0) & 0 \\ 
$ [ A_\mu \times T_{\mu\nu} ]
	[ A_\mu \times T_{\mu\rho} ] $ & II & 0 & $-0.150$(0) & 0 \\ 
$ [ A_\mu \times T_{\mu\nu} ]
	[ A_\mu \times T_{\nu\rho} ] $ & I & 0 & $-0.083$(0) & 0 \\ 
$ [ A_\mu \times T_{\mu\nu} ]
	[ A_\mu \times T_{\nu\rho} ] $ & II & 0 & $+0.249$(0) & 0 \\ 
$ [ A_\mu \times T_{\nu\rho} ]
	[ A_\mu \times T_{\mu\nu} ] $ & I & 0 & $-0.083$(0) &  0 \\ 
$ [ A_\mu \times T_{\nu\rho} ]
	[ A_\mu \times T_{\mu\nu} ] $ & II & 0 & $+0.249$(0) &  0 \\ 
$ [ A_\mu \times T_{\nu\rho} ]
	[ A_\mu \times T_{\nu\rho} ] $ & I & 0 & $+3.354$(0) & 0 \\ 
$ [ A_\mu \times T_{\nu\rho} ]
	[ A_\mu \times T_{\nu\rho} ] $ & II & 0 & $+0.804$(0) &  0 \\ 
$ [ A_\mu \times T_{\nu\rho} ]
	[ A_\mu \times T_{\nu\eta} ] $ & I & 0 & $+0.050$(0) &  0 \\ 
$ [ A_\mu \times T_{\nu\rho} ]
	[ A_\mu \times T_{\nu\eta} ] $ & II & 0 & $-0.150$(0) &  0 \\ 
$ [ A_\mu \times P ][ A_\mu \times P ] $ & I & $-16$ & $+14.269$(2) & $-2c_{TI}$ \\ 
$ [ A_\mu \times P ][ A_\mu \times P ] $ & II & 0 & $+0.725$(1) & 0 \\ 
\hline
$ [ P \times V_\mu ][ P \times V_\mu ] $ & I & 0 & $+24.465$(2) & $-c_{TI}$ \\ 
$ [ P \times V_\mu ][ P \times V_\mu ] $ & II & 0 & $-0.997$(1) &  0 \\ 
$ [ P \times V_\mu ][ P \times V_\nu ] $ & I & 0 & $-0.166$(0) &  0 \\ 
$ [ P \times V_\mu ][ P \times V_\nu ] $ & II & 0 & $+0.499$(0) &  0  \\ 
$ [ P \times A_\mu ][ P \times A_\mu ] $ & I & 0 & $-1.952$(0) &  0  \\ 
$ [ P \times A_\mu ][ P \times A_\mu ] $ & II & 0 & $+0.685$(0) &  0 \\ 
$ [ P \times A_\mu ][ P \times A_\nu ] $ & I & 0 & $-0.079$(0) &  0 \\ 
$ [ P \times A_\mu ][ P \times A_\nu ] $ & II & 0 & $+0.236$(0) &  0 \\ 
\hline
\end{tabular}
\end{center}
\caption{Renormalization constants of a color one trace four-fermion
operator at one loop,  $({\cal O}^{Latt}_2)_{I}$ defined in Eq. 
(\ref{eq:latt-basis}). 
%
%
The $\hat{\Gamma}_{ij}$ matrix represents an anomalous dimension
defined in Eq.~(\ref{eq:ff-latt-Z}) and its values are given in
Eq.~(\ref{eq:ff-anomalous}).
The $\hat{C}^{Latt}_{ij}$ matrix is defined in
Eq.~(\ref{eq:ff-latt-Z}).
The $M_{ij}$ matrix represents the tadpole-improvement
defined in Eq.~(\ref{eq:ff-TI-M}).
$c_{TI} \equiv (4\pi)^2 / 12$.
All the greek indices are summed under the condition of
$\mu \ne \nu \ne \rho \ne \eta$.
}
\label{tab:ff-op-2-I-2}
\end{table}

\clearpage

%
%
%
\begin{table}[h]
\begin{center}
\begin{tabular}{ | l | c || c | r | r | }
\hline 
${\cal O}^{Latt(0)}_j $ 	& color trace 	& $\hat{\Gamma}_{ij}$
 	& $\hat{C}^{Latt}_{ij}$ 	& $ N_{ij} $ \\ \hline
\hline
$ [ S \times V_\mu ][ S \times V_\mu ] $ & I & 0 & $+5.581$(1) & 0 \\ 
$ [ S \times V_\mu ][ S \times V_\mu ] $ & II & 0 & $-1.860$(0) & 0 \\ 
$ [ S \times V_\mu ][ S \times V_\nu ] $ & I & 0 & $-1.114$(0) & 0 \\ 
$ [ S \times V_\mu ][ S \times V_\nu ] $ & II & 0 & $+0.371$(0) & 0 \\ 
$ [ S \times A_\mu ][ S \times A_\mu ] $ & I & 0 & $-1.311$(0) & 0 \\ 
$ [ S \times A_\mu ][ S \times A_\mu ] $ & II & 0 & $+0.437$(0) & 0 \\ 
$ [ S \times A_\mu ][ S \times A_\nu ] $ & I & 0 & $-0.236$(0) & 0 \\ 
$ [ S \times A_\mu ][ S \times A_\nu ] $ & II & 0 & $+0.079$(0) & 0 \\ 
\hline
$ [ V_\mu \times S ][ V_\mu \times S ] $ & I & 0 & $+1.107$(0) & 0 \\ 
$ [ V_\mu \times S ][ V_\mu \times S ] $ & II & 0 & $-0.369$(0) & 0 \\ 
$ [ V_\mu \times T_{\mu\nu} ]
	[ V_\mu \times T_{\mu\nu} ] $ & I & 0 & $-1.106$(0) & 0 \\ 
$ [ V_\mu \times T_{\mu\nu} ]
	[ V_\mu \times T_{\mu\nu} ] $ & II & 0 & $+0.369$(0) & 0 \\ 
$ [ V_\mu \times T_{\mu\nu} ]
	[ V_\mu \times T_{\mu\rho} ] $ & I & 0 & $+0.150$(0) & 0 \\ 
$ [ V_\mu \times T_{\mu\nu} ]
	[ V_\mu \times T_{\mu\rho} ] $ & II & 0 & $-0.050$(0) & 0 \\ 
$ [ V_\mu \times T_{\mu\nu} ]
	[ V_\mu \times T_{\nu\rho} ] $ & I & 0 & $+0.403$(0) & 0 \\ 
$ [ V_\mu \times T_{\mu\nu} ]
	[ V_\mu \times T_{\nu\rho} ] $ & II & 0 & $-0.134$(0) & 0 \\ 
$ [ V_\mu \times T_{\nu\rho} ]
	[ V_\mu \times T_{\mu\nu} ] $ & I & 0 & $+0.403$(0) & 0 \\ 
$ [ V_\mu \times T_{\nu\rho} ]
	[ V_\mu \times T_{\mu\nu} ] $ & II & 0 & $-0.134$(0) & 0 \\ 
$ [ V_\mu \times T_{\nu\rho} ]
	[ V_\mu \times T_{\nu\rho} ] $ & I & 0 & $-2.776$(0) & 0 \\ 
$ [ V_\mu \times T_{\nu\rho} ]
	[ V_\mu \times T_{\nu\rho} ] $ & II & 0 & $+0.925$(0) & 0 \\ 
$ [ V_\mu \times T_{\nu\rho} ]
	[ V_1 \times T_{\nu\eta} ] $ & I & 0 & $+0.150$(0) &  0 \\ 
$ [ V_\mu \times T_{\nu\rho} ]
	[ V_1 \times T_{\nu\eta} ] $ & II & 0 & $-0.050$(0) &  0 \\ 
$ [ V_\mu \times P ]
	[ V_\mu \times P ] $ & I & $+6$ & $+4.500$(0) & 0 \\ 
$ [ V_\mu \times P ]
	[ V_\mu \times P ] $ & II & $-2$ & $-61.523$(7) & $+4c_{TI}$ \\ 
\hline
$ [ T_{\mu\nu} \times V_\mu ]
	[ T_{\mu\nu} \times V_\mu ] $ & I & 0 & $-2.767$(0) & 0 \\ 
$ [ T_{\mu\nu} \times V_\mu ]
	[ T_{\mu\nu} \times V_\mu ] $ & II & 0 & $+0.922$(0) & 0 \\ 
$ [ T_{\mu\nu} \times V_\mu ]
	[ T_{\mu\nu} \times V_\nu ] $ & I & 0 & $-0.416$(0) & 0 \\ 
$ [ T_{\mu\nu} \times V_\mu ]
	[ T_{\mu\nu} \times V_\nu ] $ & II & 0 & $+0.139$(0) & 0 \\ 
$ [ T_{\mu\nu} \times V_\mu ]
	[ T_{\mu\nu} \times V_\rho ] $ & I & 0 & $+0.765$(0) & 0 \\ 
$ [ T_{\mu\nu} \times V_\mu ]
	[ T_{\mu\nu} \times V_\rho ] $ & II & 0 & $-0.255$(0) & 0 \\ 
$ [ T_{\mu\nu} \times V_\rho ]
	[ T_{\mu\nu} \times V_\mu ] $ & I & 0 & $+0.765$(0) & 0 \\ 
$ [ T_{\mu\nu} \times V_\rho ]
	[ T_{\mu\nu} \times V_\mu ] $ & II & 0 & $-0.255$(0) & 0 \\ 
$ [ T_{\mu\nu} \times V_\rho ]
	[ T_{\mu\nu} \times V_\rho ] $ & I & 0 & $-5.968$(1) & 0 \\ 
$ [ T_{\mu\nu} \times V_\rho ]
	[ T_{\mu\nu} \times V_\rho ] $ & II & 0 & $+1.989$(0) & 0 \\ 
$ [ T_{\mu\nu} \times V_\rho ]
	[ T_{\mu\nu} \times V_\eta ] $ & I & 0 & $+1.946$(0) &  0 \\ 
$ [ T_{\mu\nu} \times V_\rho ]
	[ T_{\mu\nu} \times V_\eta ] $ & II & 0 & $-0.649$(0) & 0 \\ 
\hline
\end{tabular}
\end{center}
\caption{Renormalization constants of a color two trace four-fermion
operator at one loop, $({\cal O}^{Latt}_2)_{II}$ defined in Eq.
(\ref{eq:latt-basis}).
%
%
The $\hat{\Gamma}_{ij}$ matrix represents an anomalous dimension
defined in Eq.~(\ref{eq:ff-latt-Z}) and its values are given in
Eq.~(\ref{eq:ff-anomalous}).
The $\hat{C}^{Latt}_{ij}$ matrix is defined in
Eq.~(\ref{eq:ff-latt-Z}).
The $N_{ij}$ matrix represents the tadpole-improvement
defined in Eq.~(\ref{eq:ff-TI-N}).
$c_{TI} \equiv (4\pi)^2 / 12$.
All the greek indices are summed under the condition of
$\mu \ne \nu \ne \rho \ne \eta$.
}
\label{tab:ff-op-2-II-1}
\end{table}
\begin{table}[h]
\begin{center}
\begin{tabular}{ | l | c || c | r | r | }
\hline
${\cal O}^{Latt(0)}_j $ 	& color trace 	& $\hat{\Gamma}_{ij}$
 	& $\hat{C}^{Latt}_{ij}$ 	& $ N_{ij} $ \\ \hline
\hline 
$ [ T_{\mu\nu} \times A_\mu ]
	[ T_{\mu\nu} \times A_\mu ] $ & I & 0 & $+0.772$(0) & 0 \\ 
$ [ T_{\mu\nu} \times A_\mu ]
	[ T_{\mu\nu} \times A_\mu ] $ & II & 0 & $-0.257$(0) & 0 \\ 
$ [ T_{\mu\nu} \times A_\mu ]
	[ T_{\mu\nu} \times A_\nu ] $ & I & 0 & $+0.045$(0) & 0 \\ 
$ [ T_{\mu\nu} \times A_\mu ]
	[ T_{\mu\nu} \times A_\nu ] $ & II & 0 & $-0.015$(0) & 0 \\ 
$ [ T_{\mu\nu} \times A_\rho ]
	[ T_{\mu\nu} \times A_\rho ] $ & I & 0 & $+1.705$(0) & 0 \\ 
$ [ T_{\mu\nu} \times A_\rho ]
	[ T_{\mu\nu} \times A_\rho ] $ & II & 0 & $-0.568$(0) & 0 \\ 
$ [ T_{\mu\nu} \times A_\rho ]
	[ T_{\mu\nu} \times A_\eta ] $ & I & 0 & $+0.529$(0) & 0 \\ 
$ [ T_{\mu\nu} \times A_\rho ]
	[ T_{\mu\nu} \times A_\eta ] $ & II & 0 & $-0.176$(0) & 0 \\ 
\hline
$ [ A_\mu \times S ][ A_\mu \times S ] $ & I & 0 & $-0.478$(0) & 0 \\ 
$ [ A_\mu \times S ][ A_\mu \times S ] $ & II & 0 & $+0.159$(0) & 0 \\ 
$ [ A_\mu \times T_{\mu\nu} ]
	[ A_\mu \times T_{\mu\nu} ] $ & I & 0 & $+1.106$(0) & 0 \\ 
$ [ A_\mu \times T_{\mu\nu} ]
	[ A_\mu \times T_{\mu\nu} ] $ & II & 0 & $-0.369$(0) & 0 \\ 
$ [ A_\mu \times T_{\mu\nu} ]
	[ A_\mu \times T_{\mu\rho} ] $ & I & 0 & $-0.150$(0) & 0 \\ 
$ [ A_\mu \times T_{\mu\nu} ]
	[ A_\mu \times T_{\mu\rho} ] $ & II & 0 & $+0.050$(0) & 0 \\ 
$ [ A_\mu \times T_{\nu\rho} ]
	[ A_\mu \times T_{\nu\rho} ] $ & I & 0 & $+2.776$(0) & 0 \\ 
$ [ A_\mu \times T_{\nu\rho} ]
	[ A_\mu \times T_{\nu\rho} ] $ & II & 0 & $-0.925$(0) & 0 \\ 
$ [ A_\mu \times T_{\nu\rho} ]
	[ A_\mu \times T_{\nu\eta} ] $ & I & 0 & $+0.656$(0) & 0 \\ 
$ [ A_\mu \times T_{\nu\rho} ]
	[ A_\mu \times T_{\nu\eta} ] $ & II & 0 & $-0.219$(0) & 0 \\ 
$ [ A_\mu \times P ][ A_\mu \times P ] $ & I & $-6$ & $-4.500$(0) & 0 \\ 
$ [ A_\mu \times P ][ A_\mu \times P ] $ & II & $+2$ & $+1.498$(2) & 0 \\ 
\hline
$ [ P \times V_\mu ][ P \times V_\mu ] $ & I & 0 & $+3.153$(1) & 0 \\ 
$ [ P \times V_\mu ][ P \times V_\mu ] $ & II & 0 & $-1.051$(0) & 0 \\ 
$ [ P \times V_\mu ][ P \times V_\nu ] $ & I & 0 & $-0.416$(0) & 0 \\ 
$ [ P \times V_\mu ][ P \times V_\nu ] $ & II & 0 & $+0.139$(0) & 0 \\ 
$ [ P \times A_\mu ][ P \times A_\mu ] $ & I & 0 & $-0.377$(0) & 0 \\ 
$ [ P \times A_\mu ][ P \times A_\mu ] $ & II & 0 & $+0.126$(0) & 0 \\ 
$ [ P \times A_\mu ][ P \times A_\nu ] $ & I & 0 & $+0.045$(0) & 0 \\ 
$ [ P \times A_\mu ][ P \times A_\nu ] $ & II & 0 & $-0.015$(0) & 0 \\ 
\hline
\end{tabular}
\end{center}
\caption{Renormalization constants of a color two trace four-fermion
operator at one loop, $({\cal O}^{Latt}_2)_{II}$ defined in Eq.
(\ref{eq:latt-basis}).
%
%
The $\hat{\Gamma}_{ij}$ matrix represents an anomalous dimension
defined in Eq.~(\ref{eq:ff-latt-Z}) and its values are given in
Eq.~(\ref{eq:ff-anomalous}).
The $\hat{C}^{Latt}_{ij}$ matrix is defined in
Eq.~(\ref{eq:ff-latt-Z}).
The $N_{ij}$ matrix represents the tadpole-improvement
defined in Eq.~(\ref{eq:ff-TI-N}).
$c_{TI} \equiv (4\pi)^2 / 12$.
All the greek indices are summed under the condition of
$\mu \ne \nu \ne \rho \ne \eta$.
}
\label{tab:ff-op-2-II-2}
\end{table}

\clearpage

%
%
%
\begin{table}[h]
\begin{center}
\begin{tabular}{ | l | c || c | r | r | }
\hline 
${\cal O}^{Latt(0)}_j $ 	& color trace 	& $\hat{\Gamma}_{ij}$
 	& $\hat{C}^{Latt}_{ij}$ 	& $ M_{ij} $ \\ \hline
\hline
$ [ S \times S ][ S \times S ] $ & I & 0 & $-0.155$(0) & 0 \\ 
$ [ S \times S ][ S \times S ] $ & II & 0 & $-0.507$(0) & 0 \\ 
$ [ S \times T_{\mu\nu} ]
	[ S \times T_{\mu\nu} ] $ & I & 0 & $-1.068$(0) & 0 \\ 
$ [ S \times T_{\mu\nu} ]
	[ S \times T_{\mu\nu} ] $ & II & 0 & $-0.458$(0) & 0 \\ 
$ [ S \times T_{\mu\nu} ]
	[ S \times T_{\mu\rho} ] $ & I & 0 & $-0.349$(0) & 0 \\ 
$ [ S \times T_{\mu\nu} ]
	[ S \times T_{\mu\rho} ] $ & II & 0 & $-0.150$(0) & 0 \\ 
$ [ S \times P ]
	[ S \times P ] $ & I & $2\times(+2)$ & 
	$2\times(+25.186(2))$ & $2\times(-2c_{TI})$ \\ 
$ [ S \times P ]
	[ S \times P ] $ & II & $2\times(-6)$ 
	& $2\times(+3.972(1))$ & 0 \\ 
\hline
$ [ V_\mu \times V_\mu ]
	[ V_\mu \times V_\mu ] $ & I & 0 & $+33.748$(1) & $-2c_{TI}$ \\ 
$ [ V_\mu \times V_\mu ]
	[ V_\mu \times V_\mu ] $ & II & 0 & $-2.258$(1) & 0 \\ 
$ [ V_\mu \times V_\nu ]
	[ V_\mu \times V_\nu ] $ & I & 0 & $-0.581$(1) & 0 \\ 
$ [ V_\mu \times V_\nu ]
	[ V_\mu \times V_\nu ] $ & II & 0 & $+0.194$(0) & 0 \\ 
$ [ V_\mu \times V_\nu ]
	[ V_\mu \times V_\rho ] $ & I & 0 & $+1.249$(0) & 0 \\ 
$ [ V_\mu \times V_\nu ]
	[ V_\mu \times V_\rho ] $ & II & 0 & $-0.416$(0) & 0 \\ 
\hline
$ [ V_\mu \times A_\mu ]
	[ V_\mu \times A_\mu ] $ & I & 0 & $-0.402$(0) & 0 \\ 
$ [ V_\mu \times A_\mu ]
	[ V_\mu \times A_\mu ] $ & II & 0 & $+0.776$(0) & 0 \\ 
$ [ V_\mu \times A_\nu ]
	[ V_\mu \times A_\nu ] $ & I & 0 & $+0.461$(0) & 0 \\ 
$ [ V_\mu \times A_\nu ]
	[ V_\mu \times A_\nu ] $ & II & 0 & $+0.198$(0) & 0 \\ 
$ [ V_\mu \times A_\mu ]
	[ V_\mu \times A_\nu ] $ & I & 0 & $-0.064$(0) & 0 \\ 
$ [ V_\mu \times A_\mu ]
	[ V_\mu \times A_\nu ] $ & II & 0 & $+0.191$(0) & 0 \\ 
$ [ V_\mu \times A_\nu ]
	[ V_\mu \times A_\mu ] $ & I & 0 & $-0.064$(0) & 0 \\ 
$ [ V_\mu \times A_\nu ]
	[ V_\mu \times A_\mu ] $ & II & 0 & $+0.191$(0) & 0 \\ 
$ [ V_\mu \times A_\nu ]
	[ V_\mu \times A_\rho ] $ & I & 0 & $+0.104$(0) & 0 \\ 
$ [ V_\mu \times A_\nu ]
	[ V_\mu \times A_\rho ] $ & II & 0 & $+0.045$(0) & 0 \\ 
\hline
$ [ T_{\mu\nu} \times T_{\mu\nu} ]
	[ T_{\mu\nu} \times T_{\mu\nu} ] $ & I & 0 & $+0.815$(0) & 0 \\ 
$ [ T_{\mu\nu} \times T_{\mu\nu} ]
	[ T_{\mu\nu} \times T_{\mu\nu} ] $ & II & 0 & $-2.065$(0) & 0 \\ 
$ [ T_{\mu\nu} \times T_{\mu\rho} ]
  	[ T_{\mu\nu} \times T_{\mu\eta} ] $ & I & 0 & $+0.449$(0) & 0 \\ 
$ [ T_{\mu\nu} \times T_{\mu\rho} ]
	[ T_{\mu\nu} \times T_{\mu\eta} ] $ & II & 0 & $-0.150$(0) & 0 \\ 
$ [ T_{\mu\nu} \times T_{\mu\rho} ]
	[ T_{\mu\nu} \times T_{\nu\rho} ] $ & I & 0 & $-0.449$(0) & 0 \\ 
$ [ T_{\mu\nu} \times T_{\mu\rho} ]
	[ T_{\mu\nu} \times T_{\nu\rho} ] $ & II & 0 & $+0.150$(0) & 0 \\ 
$ [ T_{\mu\nu} \times T_{\mu\rho} ]
	[ T_{\mu\nu} \times T_{\rho\eta} ] $ & I & 0 & $+0.102$(0) & 0 \\ 
$ [ T_{\mu\nu} \times T_{\mu\rho} ]
	[ T_{\mu\nu} \times T_{\rho\eta} ] $ & II & 0 & $-0.307$(0) & 0 \\ 
$ [ T_{\mu\nu} \times T_{\rho\eta} ]
	[ T_{\mu\nu} \times T_{\mu\rho} ] $ & I & 0 & $+0.102$(0) & 0 \\ 
$ [ T_{\mu\nu} \times T_{\rho\eta} ]
	[ T_{\mu\nu} \times T_{\mu\rho} ] $ & II & 0 & $-0.307$(0) & 0 \\ 
$ [ T_{\mu\nu} \times T_{\rho\eta} ]
	[ T_{\mu\nu} \times T_{\rho\eta} ] $ & I & 0 & $-0.552$(0) & 0 \\ 
$ [ T_{\mu\nu} \times T_{\rho\eta} ]
	[ T_{\mu\nu} \times T_{\rho\eta} ] $ & II & 0 & $+1.275$(0) & 0 \\ 
\hline
\end{tabular}
\end{center}
\caption{ Renormalization constants of a color one trace four-fermion
 operator at one loop, $({\cal O}^{Latt}_3)_{I}$ defined in Eq. 
(\ref{eq:latt-basis}).
%
%
The $\hat{\Gamma}_{ij}$ matrix represents an anomalous dimension
defined in Eq.~(\ref{eq:ff-latt-Z}) and its values are given in
Eq.~(\ref{eq:ff-anomalous}).
The $\hat{C}^{Latt}_{ij}$ matrix is defined in
Eq.~(\ref{eq:ff-latt-Z}).
The $M_{ij}$ matrix represents the tadpole-improvement
defined in Eq.~(\ref{eq:ff-TI-M}).
$c_{TI} \equiv (4\pi)^2 / 12$.
All the greek indices are summed under the condition of
$\mu \ne \nu \ne \rho \ne \eta$.
}
\label{tab:ff-op-3-I-1}
\end{table}
\begin{table}[h]
\begin{center}
\begin{tabular}{ | l | c || c | r | r | }
\hline
${\cal O}^{Latt(0)}_j $ 	& color trace 	& $\hat{\Gamma}_{ij}$
 	& $\hat{C}^{Latt}_{ij}$ 	& $ M_{ij} $ \\ \hline
\hline 
$ [ A_\mu \times V_\mu ]
	[ A_\mu \times V_\mu ] $ & I & 0 & $+32.138$(1) & $-2c_{TI}$ \\ 
$ [ A_\mu \times V_\mu ]
	[ A_\mu \times V_\mu ] $ & II & 0 & $+2.572$(1) & 0 \\ 
$ [ A_\mu \times V_\nu ]
	[ A_\mu \times V_\nu ] $ & I & 0 & $-0.452$(1) & 0 \\ 
$ [ A_\mu \times V_\nu ]
	[ A_\mu \times V_\nu ] $ & II & 0 & $-0.194$(0) & 0 \\ 
$ [ A_\mu \times V_\mu ]
	[ A_\mu \times V_\nu ] $ & I & 0 & $+0.305$(0) & 0 \\ 
$ [ A_\mu \times V_\mu ]
	[ A_\mu \times V_\nu ] $ & II & 0 & $-0.915$(0) & 0 \\ 
$ [ A_\mu \times V_\nu ]
	[ A_\mu \times V_\mu ] $ & I & 0 & $+0.305$(0) & 0 \\ 
$ [ A_\mu \times V_\nu ]
	[ A_\mu \times V_\mu ] $ & II & 0 & $-0.915$(0) & 0 \\ 
$ [ A_\mu \times V_\nu ]
	[ A_\mu \times V_\rho ] $ & I & 0 & $+0.971$(0) & 0 \\ 
$ [ A_\mu \times V_\nu ]
	[ A_\mu \times V_\rho ] $ & II & 0 & $+0.416$(0) & 0 \\ 
\hline
$ [ A_\mu \times A_\mu ]
	[ A_\mu \times A_\mu ] $ & I & 0 & $+0.220$(0) & 0 \\ 
$ [ A_\mu \times A_\mu ]
	[ A_\mu \times A_\mu ] $ & II & 0 & $-1.090$(0) & 0 \\ 
$ [ A_\mu \times A_\nu ]
	[ A_\mu \times A_\nu ] $ & I & 0 & $+0.593$(0) & 0 \\ 
$ [ A_\mu \times A_\nu ]
	[ A_\mu \times A_\nu ] $ & II & 0 & $-0.198$(0) & 0 \\ 
$ [ A_\mu \times A_\nu ]
	[ A_\mu \times A_\rho ] $ & I & 0 & $+0.134$(0) & 0 \\ 
$ [ A_\mu \times A_\nu ]
	[ A_\mu \times A_\rho ] $ & II & 0 & $-0.045$(0) & 0 \\ 
\hline
$ [ P \times S ][ P \times S ] $ & I & 0 & $+0.155$(0) & 0 \\ 
$ [ P \times S ][ P \times S ] $ & II & 0 & $+0.507$(0) & 0 \\ 
$ [ P \times T_{\mu\nu} ]
	[ P \times T_{\mu\nu} ] $ & I & 0 & $+1.068$(0) & 0 \\ 
$ [ P \times T_{\mu\nu} ]
	[ P \times T_{\mu\nu} ] $ & II & 0 & $+0.458$(0) & 0 \\ 
$ [ P \times T_{\mu\nu} ]
	[ P \times T_{\mu\rho} ] $ & I & 0 & $+0.349$(0) & 0 \\ 
$ [ P \times T_{\mu\nu} ]
	[ P \times T_{\mu\rho} ] $ & II & 0 & $+0.150$(0) & 0 \\ 
$ [ P \times P ][ P \times P ] $ & I & $2\times(-2)$ & 
	$2\times(-39.872(2))$ &  $2\times(+2c_{TI})$ \\ 
$ [ P \times P ][ P \times P ] $ & II & $2\times(+6)$ & 
	$2\times(+40.087(1))$ & 0 \\ 
\hline
\end{tabular}
\end{center}
\caption{ Renormalization constants of a color one trace four fermion
 operator at one loop, $({\cal O}^{Latt}_3)_{I}$ defined in Eq. 
(\ref{eq:latt-basis}).
%
%
The $\hat{\Gamma}_{ij}$ matrix represents an anomalous dimension
defined in Eq.~(\ref{eq:ff-latt-Z}) and its values are given in
Eq.~(\ref{eq:ff-anomalous}).
The $\hat{C}^{Latt}_{ij}$ matrix is defined in
Eq.~(\ref{eq:ff-latt-Z}).
The $M_{ij}$ matrix represents the tadpole-improvement
defined in Eq.~(\ref{eq:ff-TI-M}).
$c_{TI} \equiv (4\pi)^2 / 12$.
All the greek indices are summed under the condition of
$\mu \ne \nu \ne \rho \ne \eta$.
}
\label{tab:ff-op-3-I-2}
\end{table}

\clearpage

%
%
%
\begin{table}[h]
\begin{center}
\begin{tabular}{ | l | c || c | r | r | }
\hline 
${\cal O}^{Latt(0)}_j $ 	& color trace 	& $\hat{\Gamma}_{ij}$
 	& $\hat{C}^{Latt}_{ij}$ 	& $ N_{ij} $ \\ \hline
\hline
$ [ S \times T_{\mu\nu} ]
	[ S \times T_{\mu\nu} ] $ & I & 0 & $-0.915$(0) & 0 \\ 
$ [ S \times T_{\mu\nu} ]
	[ S \times T_{\mu\nu} ] $ & II & 0 & $+0.305$(0) & 0 \\ 
$ [ S \times T_{\mu\nu} ]
	[ S \times T_{\mu\rho} ] $ & I & 0 & $-0.299$(0) & 0 \\ 
$ [ S \times T_{\mu\nu} ]
	[ S \times T_{\mu\rho} ] $ & II & 0 & $+0.100$(0) & 0 \\ 
$ [ S \times P ]
	[ S \times P ] $ & II & $2\times(-16)$ & 
	$2\times(+95.607(4))$ & $2\times(-6 c_{TI})$  \\ 
\hline
$ [ V_\mu \times A_\mu ]
	[ V_\mu \times A_\mu ] $ & I & 0 & $+0.682$(0) & 0 \\ 
$ [ V_\mu \times A_\mu ]
	[ V_\mu \times A_\mu ] $ & II & 0 & $-0.227$(0) & 0 \\ 
$ [ V_\mu \times A_\nu ]
	[ V_\mu \times A_\nu ] $ & I & 0 & $+0.395$(0) & 0 \\ 
$ [ V_\mu \times A_\nu ]
	[ V_\mu \times A_\nu ] $ & II & 0 & $-0.132$(0) & 0 \\ 
$ [ V_\mu \times A_\mu ]
	[ V_\mu \times A_\nu ] $ & I & 0 & $+0.191$(0) & 0 \\ 
$ [ V_\mu \times A_\mu ]
	[ V_\mu \times A_\nu ] $ & II & 0 & $-0.064$(0) & 0 \\ 
$ [ V_\mu \times A_\nu ]
	[ V_\mu \times A_\mu ] $ & I & 0 & $+0.191$(0) & 0 \\ 
$ [ V_\mu \times A_\nu ]
	[ V_\mu \times A_\mu ] $ & II & 0 & $-0.064$(0) & 0 \\ 
$ [ V_\mu \times A_\nu ]
	[ V_\mu \times A_\rho ] $ & I & 0 & $+0.089$(0) & 0 \\ 
$ [ V_\mu \times A_\nu ]
	[ V_\mu \times A_\rho ] $ & II & 0 & $-0.030$(0) & 0 \\ 
\hline
$ [ A_\mu \times V_\mu ]
	[ A_\mu \times V_\mu ] $ & I & 0 & $+6.015$(1) & 0 \\ 
$ [ A_\mu \times V_\mu ]
	[ A_\mu \times V_\mu ] $ & II & 0 & $-2.005$(0) & 0 \\ 
$ [ A_\mu \times V_\nu ]
	[ A_\mu \times V_\nu ] $ & I & 0 & $-0.387$(1) & 0 \\ 
$ [ A_\mu \times V_\nu ]
	[ A_\mu \times V_\nu ] $ & II & 0 & $+0.129$(0) & 0 \\ 
$ [ A_\mu \times V_\mu ]
	[ A_\mu \times V_\nu ] $ & I & 0 & $-1.530$(1) & 0 \\ 
$ [ A_\mu \times V_\mu ]
	[ A_\mu \times V_\nu ] $ & II & 0 & $+0.510$(0) & 0 \\ 
$ [ A_\mu \times V_\nu ]
	[ A_\mu \times V_\mu ] $ & I & 0 & $-1.530$(1) & 0 \\ 
$ [ A_\mu \times V_\nu ]
	[ A_\mu \times V_\mu ] $ & II & 0 & $+0.510$(0) & 0 \\ 
$ [ A_\mu \times V_\nu ]
	[ A_\mu \times V_\rho ] $ & I & 0 & $+0.833$(0) & 0 \\ 
$ [ A_\mu \times V_\nu ]
	[ A_\mu \times V_\rho ] $ & II & 0 & $-0.278$(0) & 0 \\ 
\hline
$ [ P \times T_{\mu\nu} ]
	[ P \times T_{\mu\nu} ] $ & I & 0 & $+0.915$(0) & 0 \\ 
$ [ P \times T_{\mu\nu} ]
	[ P \times T_{\mu\nu} ] $ & II & 0 & $-0.305$(0) & 0 \\ 
$ [ P \times T_{\mu\nu} ]
	[ P \times T_{\mu\rho} ] $ & I & 0 & $+0.299$(0) & 0 \\ 
$ [ P \times T_{\mu\nu} ]
	[ P \times T_{\mu\rho} ] $ & II & 0 & $-0.100$(0) & 0 \\ 
$ [ P \times P ]
	[ P \times P ] $ & II & $2\times(+16)$ & 
	$2\times(+111.255(2))$ & $2\times(-2c_{TI})$  \\ 
\hline
\end{tabular}
\end{center}
\caption{Renormalization constants of a color two trace four-fermion
operator at one loop, $({\cal O}^{Latt}_3)_{II}$ defined in Eq. 
(\ref{eq:latt-basis}). 
%
%
The $\hat{\Gamma}_{ij}$ matrix represents an anomalous dimension
defined in Eq.~(\ref{eq:ff-latt-Z}) and its values are given in
Eq.~(\ref{eq:ff-anomalous}).
The $\hat{C}^{Latt}_{ij}$ matrix is defined in
Eq.~(\ref{eq:ff-latt-Z}).
The $N_{ij}$ matrix represents the tadpole-improvement
defined in Eq.~(\ref{eq:ff-TI-N}).
$c_{TI} \equiv (4\pi)^2 / 12$.
All the greek indices are summed under the condition of
$\mu \ne \nu \ne \rho \ne \eta$.
}
\label{tab:ff-op-3-II-1}
\end{table}

\clearpage

%
%
%
\begin{table}[h]
\begin{center}
\begin{tabular}{ | l | c || c | r | r | }
\hline 
${\cal O}^{Latt(0)}_j $ 	& color trace 	& $\hat{\Gamma}_{ij}$
 	& $\hat{C}^{Latt}_{ij}$ 	& $ M_{ij} $ \\ \hline
\hline
$ [ S \times S ][ S \times S ] $ & I & 0 & $-0.077$(0) & 0 \\ 
$ [ S \times S ][ S \times S ] $ & II & 0 & $-0.254$(0) & 0  \\ 
$ [ S \times T_{\mu\nu} ]
	[ S \times T_{\mu\nu} ] $ & I & 0 & $-0.977$(0) & 0 \\ 
$ [ S \times T_{\mu\nu} ]
	[ S \times T_{\mu\nu} ] $ & II & 0 & $-0.419$(0) & 0 \\ 
$ [ S \times T_{\mu\nu} ]
	[ S \times T_{\mu\rho} ] $ & I & 0 & $-0.099$(0) & 0 \\ 
$ [ S \times T_{\mu\nu} ]
	[ S \times T_{\mu\rho} ] $ & II & 0 & $-0.042$(0) & 0 \\ 
$ [ S \times P ]
	[ S \times P ] $ & I & $-2$ & $-25.186$(2) & $+2 c_{TI}$ \\ 
$ [ S \times P ]
	[ S \times P ] $ & II & $+6$ & $-3.972$(1) & 0 \\ 
\hline
$ [ V_\mu \times V_\mu ]
	[ V_\mu \times V_\mu ] $ & I & 0 & $+16.874$(1) & $-c_{TI}$ \\ 
$ [ V_\mu \times V_\mu ]
	[ V_\mu \times V_\mu ] $ & II & 0 & $-1.129$(0) & 0 \\ 
$ [ V_\mu \times V_\nu ]
	[ V_\mu \times V_\nu ] $ & I & 0 & $+2.864$(0) & 0 \\ 
$ [ V_\mu \times V_\nu ]
	[ V_\mu \times V_\nu ] $ & II & 0 & $-0.955$(0) & 0 \\ 
\hline
$ [ V_\mu \times A_\mu ]
	[ V_\mu \times A_\mu ] $ & I & 0 & $+0.201$(0) & 0 \\ 
$ [ V_\mu \times A_\mu ]
	[ V_\mu \times A_\mu ] $ & II & 0 & $-0.388$(0) & 0 \\ 
$ [ V_\mu \times A_\mu ]
	[ V_\mu \times A_\nu ] $ & I & 0 & $+0.032$(0) & 0 \\ 
$ [ V_\mu \times A_\mu ]
	[ V_\mu \times A_\nu ] $ & II & 0 & $-0.096$(0) & 0 \\ 
$ [ V_\mu \times A_\nu ]
	[ V_\mu \times A_\mu ] $ & I & 0 & $+0.032$(0) & 0 \\ 
$ [ V_\mu \times A_\nu ]
	[ V_\mu \times A_\mu ] $ & II & 0 & $-0.096$(0) & 0 \\ 
$ [ V_\mu \times A_\nu ]
	[ V_\mu \times A_\nu ] $ & I & 0 & $-0.524$(0) & 0 \\ 
$ [ V_\mu \times A_\nu ]
	[ V_\mu \times A_\nu ] $ & II & 0 & $-0.224$(0) & 0 \\ 
\hline
$ [ T_{\mu\nu} \times S ]
	[ T_{\mu\nu} \times S ] $ & I & 0 & $-0.222$(0) & 0 \\ 
$ [ T_{\mu\nu} \times S ]
	[ T_{\mu\nu} \times S ] $ & II & 0 & $-0.095$(0) & 0 \\ 
$ [ T_{\mu\nu} \times T_{\mu\nu} ]
	[ T_{\mu\nu} \times T_{\mu\nu} ] $ & I & 0 & $+0.407$(0) & 0 \\ 
$ [ T_{\mu\nu} \times T_{\mu\nu} ]
	[ T_{\mu\nu} \times T_{\mu\nu} ] $ & II & 0 & $-1.032$(0) & 0 \\ 
$ [ T_{\mu\nu} \times T_{\mu\rho} ]
	[ T_{\mu\nu} \times T_{\mu\rho} ] $ & I & 0 & $+1.374$(0) & 0 \\ 
$ [ T_{\mu\nu} \times T_{\mu\rho} ]
	[ T_{\mu\nu} \times T_{\mu\rho} ] $ & II & 0 & $-0.458$(0) & 0 \\ 
$ [ T_{\mu\nu} \times T_{\mu\rho} ]
	[ T_{\mu\nu} \times T_{\nu\rho} ] $ & I & 0 & $-0.127$(0) & 0 \\ 
$ [ T_{\mu\nu} \times T_{\mu\rho} ]
	[ T_{\mu\nu} \times T_{\nu\rho} ] $ & II & 0 & $+0.042$(0) & 0 \\ 
$ [ T_{\mu\nu} \times T_{\mu\rho} ]
	[ T_{\mu\nu} \times T_{\mu\eta} ] $ & I & 0 & $-0.127$(0) & 0 \\ 
$ [ T_{\mu\nu} \times T_{\mu\rho} ]
	[ T_{\mu\nu} \times T_{\mu\eta} ] $ & II & 0 & $+0.042$(0) & 0 \\ 
$ [ T_{\mu\nu} \times T_{\mu\rho} ]
	[ T_{\mu\nu} \times T_{\rho\eta} ] $ & I & 0 & $-0.051$(0) & 0 \\ 
$ [ T_{\mu\nu} \times T_{\mu\rho} ]
	[ T_{\mu\nu} \times T_{\rho\eta} ] $ & II & 0 & $+0.154$(0) & 0 \\ 
$ [ T_{\mu\nu} \times T_{\rho\eta} ]
	[ T_{\mu\nu} \times T_{\mu\rho} ] $ & I & 0 & $-0.051$(0) & 0 \\ 
$ [ T_{\mu\nu} \times T_{\rho\eta} ]
	[ T_{\mu\nu} \times T_{\mu\rho} ] $ & II & 0 & $+0.154$(0) & 0 \\ 
$ [ T_{\mu\nu} \times T_{\rho\eta} ]
	[ T_{\mu\nu} \times T_{\rho\eta} ] $ & I & 0 & $+0.276$(0) & 0 \\ 
$ [ T_{\mu\nu} \times T_{\rho\eta} ]
	[ T_{\mu\nu} \times T_{\rho\eta} ] $ & II & 0 & $-0.638$(0) & 0 \\ 
$ [ T_{\mu\nu} \times P ]
	[ T_{\mu\nu} \times P ] $ & I & $-14/3$ & $-3.500$(0) & 0 \\ 
$ [ T_{\mu\nu} \times P ]
	[ T_{\mu\nu} \times P ] $ & II & $-2$ & $-1.500$(0) & 0 \\ 
\hline
\end{tabular}
\end{center}
\caption{Renormalization constants of a color one trace four-fermion
operator at one loop, $({\cal O}^{Latt}_4)_{I}$ defined in Eq. 
(\ref{eq:latt-basis}).
%
%
The $\hat{\Gamma}_{ij}$ matrix represents an anomalous dimension
defined in Eq.~(\ref{eq:ff-latt-Z}) and its values are given in
Eq.~(\ref{eq:ff-anomalous}).
The $\hat{C}^{Latt}_{ij}$ matrix is defined in
Eq.~(\ref{eq:ff-latt-Z}).
The $M_{ij}$ matrix represents the tadpole-improvement
defined in Eq.~(\ref{eq:ff-TI-M}).
$c_{TI} \equiv (4\pi)^2 / 12$.
All the greek indices are summed under the condition of
$\mu \ne \nu \ne \rho \ne \eta$.
}
\label{tab:ff-op-S+P-I-1}
\end{table}
\begin{table}[h]
\begin{center}
\begin{tabular}{ | l | c || c | r | r | }
\hline
${\cal O}^{Latt(0)}_j $ 	& color trace 	& $\hat{\Gamma}_{ij}$
 	& $\hat{C}^{Latt}_{ij}$ 	& $ M_{ij} $ \\ \hline
\hline
$ [ A_\mu \times V_\mu ]
	[ A_\mu \times V_\mu ] $ & I & 0 & $-16.069$(1) & $+c_{TI}$ \\ 
$ [ A_\mu \times V_\mu ]
	[ A_\mu \times V_\mu ] $ & II & 0 & $-1.286$(0) & 0 \\ 
$ [ A_\mu \times V_\nu ]
	[ A_\mu \times V_\nu ] $ & I & 0 & $-2.228$(0) & 0 \\ 
$ [ A_\mu \times V_\nu ]
	[ A_\mu \times V_\nu ] $ & II & 0 & $-0.955$(0) & 0 \\ 
$ [ A_\mu \times V_\mu ]
	[ A_\mu \times V_\nu ] $ & I & 0 & $-0.153$(0) & 0 \\ 
$ [ A_\mu \times V_\mu ]
	[ A_\mu \times V_\nu ] $ & II & 0 & $+0.458$(0) & 0 \\ 
$ [ A_\mu \times V_\nu ]
	[ A_\mu \times V_\mu ] $ & I & 0 & $-0.153$(0) & 0 \\ 
$ [ A_\mu \times V_\nu ]
	[ A_\mu \times V_\mu ] $ & II & 0 & $+0.458$(0) & 0 \\ 
\hline
$ [ A_\mu \times A_\mu ]
	[ A_\mu \times A_\mu ] $ & I & 0 & $+0.110$(0) & 0 \\ 
$ [ A_\mu \times A_\mu ]
	[ A_\mu \times A_\mu ] $ & II & 0 & $-0.545$(0) & 0 \\ 
$ [ A_\mu \times A_\nu ]
	[ A_\mu \times A_\nu ] $ & I & 0 & $+0.673$(0) & 0 \\ 
$ [ A_\mu \times A_\nu ]
	[ A_\mu \times A_\nu ] $ & II & 0 & $-0.224$(0) & 0 \\ 
\hline
$ [ P \times S ][ P \times S ] $ & I & 0 & $-0.077$(0) & 0 \\ 
$ [ P \times S ][ P \times S ] $ & II & 0 & $-0.254$(0) & 0 \\ 
$ [ P \times T_{\mu\nu} ]
	[ P \times T_{\mu\nu} ] $ & I & 0 & $-0.977$(0) & 0 \\ 
$ [ P \times T_{\mu\nu} ]
	[ P \times T_{\mu\nu} ] $ & II & 0 & $-0.419$(0) & 0 \\ 
$ [ P \times T_{\mu\nu} ]
	[ P \times T_{\mu\rho} ] $ & I & 0 & $-0.099$(0) & 0 \\ 
$ [ P \times T_{\mu\nu} ]
	[ P \times T_{\mu\rho} ] $ & II & 0 & $-0.042$(0) & 0 \\ 
$ [ P \times P ][ P \times P ] $ & I & $-2$ & $-39.872$(2) & $+2 c_{TI}$ \\ 
$ [ P \times P ][ P \times P ] $ & II & $+6$ & $+40.087$(1) & 0 \\ 
\hline
\end{tabular}
\end{center}
\caption{Renormalization constants of a color one trace four-fermion
operator at one loop, $({\cal O}^{Latt}_4)_{I}$ defined in Eq.
(\ref{eq:latt-basis}).
%
%
The $\hat{\Gamma}_{ij}$ matrix represents an anomalous dimension
defined in Eq.~(\ref{eq:ff-latt-Z}) and its values are given in
Eq.~(\ref{eq:ff-anomalous}).
The $\hat{C}^{Latt}_{ij}$ matrix is defined in
Eq.~(\ref{eq:ff-latt-Z}).
The $M_{ij}$ matrix represents the tadpole-improvement
defined in Eq.~(\ref{eq:ff-TI-M}).
$c_{TI} \equiv (4\pi)^2 / 12$.
All the greek indices are summed under the condition of
$\mu \ne \nu \ne \rho \ne \eta$.
}
\label{tab:ff-op-S+P-I-2}
\end{table}

\clearpage

%
%
%
\begin{table}[h]
\begin{center}
\begin{tabular}{ | l | c || c | r | r | }
\hline 
${\cal O}^{Latt(0)}_j $ 	& color trace 	& $\hat{\Gamma}_{ij}$
 	& $\hat{C}^{Latt}_{ij}$ 	& $ N_{ij} $ \\ \hline
\hline
$ [ S \times T_{\mu\nu} ]
	[ S \times T_{\mu\nu} ] $ & I & 0 & $-0.837$(0) & 0 \\ 
$ [ S \times T_{\mu\nu} ]
	[ S \times T_{\mu\nu} ] $ & II & 0 & $+0.279$(0) & 0 \\ 
$ [ S \times T_{\mu\nu} ]
	[ S \times T_{\mu\rho} ] $ & I & 0 & $-0.085$(0) & 0 \\ 
$ [ S \times T_{\mu\nu} ]
	[ S \times T_{\mu\rho} ] $ & II & 0 & $+0.028$(0) & 0 \\ 
$ [ S \times P ]
	[ S \times P ] $ & II & $+16$ & $-95.607$(4) & $+6 c_{TI}$  \\ 
\hline
$ [ V_\mu \times A_\mu ]
	[ V_\mu \times A_\mu ] $ & I & 0 & $-0.341$(0) & 0 \\ 
$ [ V_\mu \times A_\mu ]
	[ V_\mu \times A_\mu ] $ & II & 0 & $+0.114$(0) & 0 \\ 
$ [ V_\mu \times A_\nu ]
	[ V_\mu \times A_\nu ] $ & I & 0 & $-0.449$(0) & 0 \\ 
$ [ V_\mu \times A_\nu ]
	[ V_\mu \times A_\nu ] $ & II & 0 & $+0.150$(0) & 0 \\ 
$ [ V_\mu \times A_\mu ]
	[ V_\mu \times A_\nu ] $ & I & 0 & $-0.096$(0) & 0 \\ 
$ [ V_\mu \times A_\mu ]
	[ V_\mu \times A_\nu ] $ & II & 0 & $+0.032$(0) & 0 \\ 
$ [ V_\mu \times A_\nu ]
	[ V_\mu \times A_\mu ] $ & I & 0 & $-0.096$(0) & 0 \\ 
$ [ V_\mu \times A_\nu ]
	[ V_\mu \times A_\mu ] $ & II & 0 & $+0.032$(0) & 0 \\ 
\hline
$ [ T_{\mu\nu} \times S ]
	[ T_{\mu\nu} \times S ] $ & I & 0 & $-0.190$(0) & 0 \\ 
$ [ T_{\mu\nu} \times S ]
	[ T_{\mu\nu} \times S ] $ & II & 0 & $+0.063$(0) & 0 \\ 
$ [ T_{\mu\nu} \times P ]
	[ T_{\mu\nu} \times P ] $ & I & $-4$ & $-3.000$(0) & 0 \\ 
$ [ T_{\mu\nu} \times P ]
	[ T_{\mu\nu} \times P ] $ & II & $+4/3$ & $+1.000$(0) & 0 \\ 
\hline
$ [ A_\mu \times V_\mu ]
	[ A_\mu \times V_\mu ] $ & I & 0 & $-3.007$(1) & 0 \\ 
$ [ A_\mu \times V_\mu ]
	[ A_\mu \times V_\mu ] $ & II & 0 & $+1.002$(0) & 0 \\ 
$ [ A_\mu \times V_\nu ]
	[ A_\mu \times V_\nu ] $ & I & 0 & $-1.909$(0) & 0 \\ 
$ [ A_\mu \times V_\nu ]
	[ A_\mu \times V_\nu ] $ & II & 0 & $+0.636$(0) & 0 \\ 
$ [ A_\mu \times V_\mu ]
	[ A_\mu \times V_\nu ] $ & I & 0 & $+0.765$(0) & 0 \\ 
$ [ A_\mu \times V_\mu ]
	[ A_\mu \times V_\nu ] $ & II & 0 & $-0.255$(0) & 0 \\ 
$ [ A_\mu \times V_\nu ]
	[ A_\mu \times V_\mu ] $ & I & 0 & $+0.765$(0) & 0 \\ 
$ [ A_\mu \times V_\nu ]
	[ A_\mu \times V_\mu ] $ & II & 0 & $-0.255$(0) & 0 \\ 
\hline
$ [ P \times T_{\mu\nu} ]
	[ P \times T_{\mu\nu} ] $ & I & 0 & $-0.837$(0) & 0 \\ 
$ [ P \times T_{\mu\nu} ]
	[ P \times T_{\mu\nu} ] $ & II & 0 & $+0.279$(0) & 0 \\ 
$ [ P \times T_{\mu\nu} ]
	[ P \times T_{\mu\rho} ] $ & I & 0 & $-0.085$(0) & 0 \\ 
$ [ P \times T_{\mu\nu} ]
	[ P \times T_{\mu\rho} ] $ & II & 0 & $+0.028$(0) & 0 \\ 
$ [ P \times P ]
	[ P \times P ] $ & II & $+16$ & $+111.255$(2) & $-2 c_{TI}$  \\ 
\hline
\end{tabular}
\end{center}
\caption{Renormalization constants of a color two trace four-fermion
operator at one loop, $({\cal O}^{Latt}_4)_{II}$ defined in Eq. 
(\ref{eq:latt-basis}).
%
%
The $\hat{\Gamma}_{ij}$ matrix represents an anomalous dimension
defined in Eq.~(\ref{eq:ff-latt-Z}) and its values are given in
Eq.~(\ref{eq:ff-anomalous}).
The $\hat{C}^{Latt}_{ij}$ matrix is defined in
Eq.~(\ref{eq:ff-latt-Z}).
The $N_{ij}$ matrix represents the tadpole-improvement
defined in Eq.~(\ref{eq:ff-TI-N}).
$c_{TI} \equiv (4\pi)^2 / 12$.
All the greek indices are summed under the condition of
$\mu \ne \nu \ne \rho \ne \eta$.
}
\label{tab:ff-op-S+P-II-1}
\end{table}

\clearpage

%
%
%
\begin{table}[h]
\begin{center}
\begin{tabular}{ | l | c || c | r | r | }
\hline 
${\cal O}^{Latt(0)}_j $ 	& color trace 	& $\hat{\Gamma}_{ij}$
 	& $\hat{C}^{Latt}_{ij}$ 	& $ M_{ij} $ \\ \hline
\hline
$ [ S \times S ]
	[ S \times S ] $ & I & 0 & $-0.294$(0) & 0 \\ 
$ [ S \times S ]
	[ S \times S ] $ & II & 0 & $-0.016$(0) & 0 \\ 
$ [ S \times T_{\mu\nu} ]
	[ S \times T_{\mu\nu} ] $ & I & 0 & $+2.000$(0) & 0 \\ 
$ [ S \times T_{\mu\nu} ]
	[ S \times T_{\mu\nu} ] $ & II & 0 & $-0.568$(0) & 0 \\ 
$ [ S \times T_{\mu\nu} ]
	[ S \times T_{\mu\rho} ] $ & I & 0 & $+0.037$(0) & 0 \\ 
$ [ S \times T_{\mu\nu} ]
	[ S \times T_{\mu\rho} ] $ & II & 0 & $-0.111$(0) & 0 \\ 
$ [ S \times P ]
	[ S \times P ] $ & I & $-6$ & $+7.343$(1) & $-c_{TI}$ \\ 
$ [ S \times P ]
	[ S \times P ] $ & II & $-6$ & $-0.264$(1) & 0 \\ 
\hline
$ [ V_\mu \times V_\mu ]
	[ V_\mu \times V_\mu ] $ & I & 0 & $-11.777$(0) & $+c_{TI}/2$ \\ 
$ [ V_\mu \times V_\mu ]
	[ V_\mu \times V_\mu ] $ & II & 0 & $-0.867$(0) & 0 \\ 
$ [ V_\mu \times V_\nu ]
	[ V_\mu \times V_\nu ] $ & I & 0 & $-11.628$(0) & $+c_{TI}/2$ \\ 
$ [ V_\mu \times V_\nu ]
	[ V_\mu \times V_\nu ] $ & II & 0 & $-1.318$(0) & 0 \\ 
$ [ V_\mu \times V_\mu ]
	[ V_\mu \times V_\nu ] $ & I & 0 & $-0.211$(0) & 0 \\ 
$ [ V_\mu \times V_\mu ]
	[ V_\mu \times V_\nu ] $ & II & 0 & $+0.632$(0) & 0 \\ 
$ [ V_\mu \times V_\nu ]
	[ V_\mu \times V_\mu ] $ & I & 0 & $-0.211$(0) & 0 \\ 
$ [ V_\mu \times V_\nu ]
	[ V_\mu \times V_\mu ] $ & II & 0 & $+0.632$(0) & 0 \\ 
\hline
$ [ V_\mu \times A_\mu ]
	[ V_\mu \times A_\mu ] $ & I & 0 & $+0.910$(0) & 0 \\ 
$ [ V_\mu \times A_\mu ]
	[ V_\mu \times A_\mu ] $ & II & 0 & $-0.143$(0) & 0 \\ 
$ [ V_\mu \times A_\nu ]
	[ V_\mu \times A_\nu ] $ & I & 0 & $+0.990$(0) & 0 \\ 
$ [ V_\mu \times A_\nu ]
	[ V_\mu \times A_\nu ] $ & II & 0 & $-0.385$(0) & 0 \\ 
$ [ V_\mu \times A_\mu ]
	[ V_\mu \times A_\nu ] $ & I & 0 & $-0.016$(0) & 0 \\ 
$ [ V_\mu \times A_\mu ]
	[ V_\mu \times A_\nu ] $ & II & 0 & $+0.048$(0) & 0 \\ 
$ [ V_\mu \times A_\nu ]
	[ V_\mu \times A_\mu ] $ & I & 0 & $-0.016$(0) & 0 \\ 
$ [ V_\mu \times A_\nu ]
	[ V_\mu \times A_\mu ] $ & II & 0 & $+0.048$(0) & 0 \\ 
\hline
\end{tabular}
\end{center}
\caption{Renormalization constants of a color one trace four-fermion
operator at one loop, $({\cal O}^{Latt}_5)_{I}$ defined in Eq. 
(\ref{eq:latt-basis}).
%
%
The $\hat{\Gamma}_{ij}$ matrix represents an anomalous dimension
defined in Eq.~(\ref{eq:ff-latt-Z}) and its values are given in
Eq.~(\ref{eq:ff-anomalous}).
The $\hat{C}^{Latt}_{ij}$ matrix is defined in
Eq.~(\ref{eq:ff-latt-Z}).
The $M_{ij}$ matrix represents the tadpole-improvement
defined in Eq.~(\ref{eq:ff-TI-M}).
$c_{TI} \equiv (4\pi)^2 / 12$.
All the greek indices are summed under the condition of
$\mu \ne \nu \ne \rho \ne \eta$.
}
\label{tab:ff-op-T-S-P-I-0}
\end{table}
\begin{table}[h]
\begin{center}
\begin{tabular}{ | l | c || c | r | r | }
\hline 
${\cal O}^{Latt(0)}_j $ 	& color trace 	& $\hat{\Gamma}_{ij}$
 	& $\hat{C}^{Latt}_{ij}$ 	& $ M_{ij} $ \\ \hline
\hline
$ [ T_{\mu\nu} \times S ]
	[ T_{\mu\nu} \times S ] $ & I & 0 & $+0.358$(0) & 0 \\ 
$ [ T_{\mu\nu} \times S ]
	[ T_{\mu\nu} \times S ] $ & II & 0 & $-0.174$(0) & 0 \\ 
$ [ T_{\mu\nu} \times T_{\mu\nu} ]
	[ T_{\mu\nu} \times T_{\mu\nu} ] $ & I & 0 & $-1.761$(0) & 0 \\ 
$ [ T_{\mu\nu} \times T_{\mu\nu} ]
	[ T_{\mu\nu} \times T_{\mu\nu} ] $ & II & 0 & $-0.151$(0) & 0 \\ 
$ [ T_{\mu\nu} \times T_{\mu\rho} ]
	[ T_{\mu\nu} \times T_{\mu\rho} ] $ & I & 0 & $-1.602$(0) & 0 \\ 
$ [ T_{\mu\nu} \times T_{\mu\rho} ]
	[ T_{\mu\nu} \times T_{\mu\rho} ] $ & II & 0 & $-0.627$(0) & 0 \\ 
$ [ T_{\mu\nu} \times T_{\mu\nu} ]
	[ T_{\mu\nu} \times T_{\mu\rho} ] $ & I & 0 & $+0.042$(0) & 0 \\ 
$ [ T_{\mu\nu} \times T_{\mu\nu} ]
	[ T_{\mu\nu} \times T_{\mu\rho} ] $ & II & 0 & $-0.125$(0) & 0 \\ 
$ [ T_{\mu\nu} \times T_{\mu\rho} ]
	[ T_{\mu\nu} \times T_{\mu\nu} ] $ & I & 0 & $+0.042$(0) & 0 \\ 
$ [ T_{\mu\nu} \times T_{\mu\rho} ]
	[ T_{\mu\nu} \times T_{\mu\nu} ] $ & II & 0 & $-0.125$(0) & 0 \\ 
$ [ T_{\mu\nu} \times T_{\mu\rho} ]
	[ T_{\mu\nu} \times T_{\nu\rho} ] $ & I & 0 & $+0.014$(0) & 0 \\ 
$ [ T_{\mu\nu} \times T_{\mu\rho} ]
	[ T_{\mu\nu} \times T_{\nu\rho} ] $ & II & 0 & $-0.042$(0) & 0 \\ 
$ [ T_{\mu\nu} \times T_{\mu\rho} ]
	[ T_{\mu\nu} \times T_{\mu\eta} ] $ & I & 0 & $+0.097$(0) & 0 \\ 
$ [ T_{\mu\nu} \times T_{\mu\rho} ]
	[ T_{\mu\nu} \times T_{\mu\eta} ] $ & II & 0 & $-0.292$(0) & 0 \\ 
$ [ T_{\mu\nu} \times T_{\mu\rho} ]
	[ T_{\mu\nu} \times T_{\rho\eta} ] $ & I & 0 & $+0.026$(0) & 0 \\ 
$ [ T_{\mu\nu} \times T_{\mu\rho} ]
	[ T_{\mu\nu} \times T_{\rho\eta} ] $ & II & 0 & $-0.077$(0) & 0 \\ 
$ [ T_{\mu\nu} \times T_{\rho\eta} ]
	[ T_{\mu\nu} \times T_{\mu\rho} ] $ & I & 0 & $+0.026$(0) & 0 \\ 
$ [ T_{\mu\nu} \times T_{\rho\eta} ]
	[ T_{\mu\nu} \times T_{\mu\rho} ] $ & II & 0 & $-0.077$(0) & 0 \\ 
$ [ T_{\mu\nu} \times T_{\rho\eta} ]
	[ T_{\mu\nu} \times T_{\rho\eta} ] $ & I & 0 & $-1.695$(0) & 0 \\ 
$ [ T_{\mu\nu} \times T_{\rho\eta} ]
	[ T_{\mu\nu} \times T_{\rho\eta} ] $ & II & 0 & $-0.349$(0) & 0 \\ 
$ [ T_{\mu\nu} \times P ]
	[ T_{\mu\nu} \times P ] $ & I & $+26/3$ & $-6.617$(1) & $+c_{TI}$ \\ 
$ [ T_{\mu\nu} \times P ]
	[ T_{\mu\nu} \times P ] $ & II & $-2$ & $-1.915$(1) & 0 \\ 
\hline
\end{tabular}
\end{center}
\caption{Renormalization constants of a color one trace four-fermion
operator at one loop, $({\cal O}^{Latt}_5)_{I}$ defined in Eq. 
(\ref{eq:latt-basis}).
%
%
The $\hat{\Gamma}_{ij}$ matrix represents an anomalous dimension
defined in Eq.~(\ref{eq:ff-latt-Z}) and its values are given in
Eq.~(\ref{eq:ff-anomalous}).
The $\hat{C}^{Latt}_{ij}$ matrix is defined in
Eq.~(\ref{eq:ff-latt-Z}).
The $M_{ij}$ matrix represents the tadpole-improvement
defined in Eq.~(\ref{eq:ff-TI-M}).
$c_{TI} \equiv (4\pi)^2 / 12$.
All the greek indices are summed under the condition of
$\mu \ne \nu \ne \rho \ne \eta$.
}
\label{tab:ff-op-T-S-P-I-1}
\end{table}
\begin{table}[h]
\begin{center}
\begin{tabular}{ | l | c || c | r | r | }
\hline
${\cal O}^{Latt(0)}_j $ 	& color trace 	& $\hat{\Gamma}_{ij}$
 	& $\hat{C}^{Latt}_{ij}$ 	& $ M_{ij} $ \\ \hline
\hline
$ [ A_\mu \times V_\mu ]
	[ A_\mu \times V_\mu ] $ & I & 0 & $+12.329$(0) & $-c_{TI}/2$ \\ 
$ [ A_\mu \times V_\mu ]
	[ A_\mu \times V_\mu ] $ & II & 0 & $-0.789$(0) & 0 \\ 
$ [ A_\mu \times V_\nu ]
	[ A_\mu \times V_\nu ] $ & I & 0 & $+12.480$(0) & $-c_{TI}/2$ \\ 
$ [ A_\mu \times V_\nu ]
	[ A_\mu \times V_\nu ] $ & II & 0 & $-1.239$(0) & 0 \\ 
$ [ A_\mu \times V_\mu ]
	[ A_\mu \times V_\nu ] $ & I & 0 & $+0.076$(0) & 0 \\ 
$ [ A_\mu \times V_\mu ]
	[ A_\mu \times V_\nu ] $ & II & 0 & $-0.229$(0) & 0 \\ 
$ [ A_\mu \times V_\nu ]
	[ A_\mu \times V_\mu ] $ & I & 0 & $+0.076$(0) & 0 \\ 
$ [ A_\mu \times V_\nu ]
	[ A_\mu \times V_\mu ] $ & II & 0 & $-0.229$(0) & 0 \\ 
$ [ A_\mu \times V_\nu ]
	[ A_\mu \times V_\rho ] $ & I & 0 & $-0.223$(0) & 0 \\ 
$ [ A_\mu \times V_\nu ]
	[ A_\mu \times V_\rho ] $ & II & 0 & $+0.669$(0) & 0 \\ 
\hline
$ [ A_\mu \times A_\mu ]
	[ A_\mu \times A_\mu ] $ & I & 0 & $-0.841$(0) & 0 \\ 
$ [ A_\mu \times A_\mu ]
	[ A_\mu \times A_\mu ] $ & II & 0 & $-0.064$(0) & 0 \\ 
$ [ A_\mu \times A_\nu ]
	[ A_\mu \times A_\nu ] $ & I & 0 & $-0.760$(0) & 0 \\ 
$ [ A_\mu \times A_\nu ]
	[ A_\mu \times A_\nu ] $ & II & 0 & $-0.306$(0) & 0 \\ 
$ [ A_\mu \times A_\mu ]
	[ A_\mu \times A_\nu ] $ & I & 0 & $+0.016$(0) & 0 \\ 
$ [ A_\mu \times A_\mu ]
	[ A_\mu \times A_\nu ] $ & II & 0 & $-0.048$(0) & 0 \\ 
$ [ A_\mu \times A_\nu ]
	[ A_\mu \times A_\mu ] $ & I & 0 & $+0.016$(0) & 0 \\ 
$ [ A_\mu \times A_\nu ]
	[ A_\mu \times A_\mu ] $ & II & 0 & $-0.048$(0) & 0 \\ 
$ [ A_\mu \times A_\nu ]
	[ A_\mu \times A_\rho ] $ & I & 0 & $+0.032$(0) & 0 \\ 
$ [ A_\mu \times A_\nu ]
	[ A_\mu \times A_\rho ] $ & II & 0 & $-0.096$(0) & 0 \\ 
\hline		           		       
$ [ P \times S ][ P \times S ] $ & I & 0 & $-0.294$(0) & 0 \\ 
$ [ P \times S ][ P \times S ] $ & II & 0 & $-0.016$(0) & 0 \\ 
$ [ P \times T_{\mu\nu} ]
	[ P \times T_{\mu\nu} ] $ & I & 0 & $+2.066$(0) & 0 \\ 
$ [ P \times T_{\mu\nu} ]
	[ P \times T_{\mu\nu} ] $ & II & 0 & $-0.765$(0) & 0 \\ 
$ [ P \times T_{\mu\nu} ]
	[ P \times T_{\mu\rho} ] $ & I & 0 & $-0.014$(0) & 0 \\ 
$ [ P \times T_{\mu\nu} ]
	[ P \times T_{\mu\rho} ] $ & II & 0 & $+0.042$(0) & 0 \\ 
$ [ P \times P ]
	[ P \times P ] $ & I & $-6$ & $+14.686$(1) & $-c_{TI}$ \\ 
$ [ P \times P ]
	[ P \times P ] $ & II & $-6$ & $-22.293$(1) & 0 \\ 
\hline
\end{tabular}
\end{center}
\caption{Renormalization constants of a color one trace four-fermion
operator at one loop, $({\cal O}^{Latt}_5)_{I}$ defined in Eq. 
(\ref{eq:latt-basis}).
%
%
The $\hat{\Gamma}_{ij}$ matrix represents an anomalous dimension
defined in Eq.~(\ref{eq:ff-latt-Z}) and its values are given in
Eq.~(\ref{eq:ff-anomalous}).
The $\hat{C}^{Latt}_{ij}$ matrix is defined in
Eq.~(\ref{eq:ff-latt-Z}).
The $M_{ij}$ matrix represents the tadpole-improvement
defined in Eq.~(\ref{eq:ff-TI-M}).
$c_{TI} \equiv (4\pi)^2 / 12$.
All the greek indices are summed under the condition of
$\mu \ne \nu \ne \rho \ne \eta$.
}
\label{tab:ff-op-T-S-P-I-2}
\end{table}

\clearpage

%
%
%
\begin{table}[h]
\begin{center}
\begin{tabular}{ | l | c || c | r | r | }
\hline 
${\cal O}^{Latt(0)}_j $ 	& color trace 	& $\hat{\Gamma}_{ij}$
 	& $\hat{C}^{Latt}_{ij}$ 	& $ N_{ij} $ \\ \hline
\hline
$ [ S \times S ]
	[ S \times S ] $ & I & 0 & $-0.286$(0) & 0 \\ 
$ [ S \times S ]
	[ S \times S ] $ & II & 0 & $+0.095$(0) & 0 \\ 
$ [ S \times T_{\mu\nu} ]
	[ S \times T_{\mu\nu} ] $ & I & 0 & $+0.419$(0) & 0 \\ 
$ [ S \times T_{\mu\nu} ]
	[ S \times T_{\mu\nu} ] $ & II & 0 & $-0.140$(0) & 0 \\ 
$ [ S \times T_{\mu\nu} ]
	[ S \times T_{\mu\rho} ] $ & I & 0 & $+0.042$(0) & 0 \\ 
$ [ S \times T_{\mu\nu} ]
	[ S \times T_{\mu\rho} ] $ & II & 0 & $-0.014$(0) & 0 \\ 
$ [ S \times P ]
	[ S \times P ] $ & I & $-6$ & $-4.500$(0) & 0 \\ 
$ [ S \times P ]
	[ S \times P ] $ & II & $-6$ & $+49.304$(2) & $-3 c_{TI}$ \\ 
\hline
$ [ V_\mu \times V_\mu ]
	[ V_\mu \times V_\mu ] $ & I & 0 & $-2.863$(0) & 0 \\ 
$ [ V_\mu \times V_\mu ]
	[ V_\mu \times V_\mu ] $ & II & 0 & $+0.954$(0) & 0 \\ 
$ [ V_\mu \times V_\nu ]
	[ V_\mu \times V_\nu ] $ & I & 0 & $-3.413$(0) & 0 \\ 
$ [ V_\mu \times V_\nu ]
	[ V_\mu \times V_\nu ] $ & II & 0 & $+1.138$(0) & 0 \\ 
$ [ V_\mu \times V_\mu ]
	[ V_\mu \times V_\nu ] $ & I & 0 & $+0.382$(0) & 0 \\ 
$ [ V_\mu \times V_\mu ]
	[ V_\mu \times V_\nu ] $ & II & 0 & $-0.127$(0) & 0 \\ 
$ [ V_\mu \times V_\nu ]
	[ V_\mu \times V_\mu ] $ & I & 0 & $+0.382$(0) & 0 \\ 
$ [ V_\mu \times V_\nu ]
	[ V_\mu \times V_\mu ] $ & II & 0 & $-0.127$(0) & 0 \\ 
$ [ V_\mu \times V_\nu ]
	[ V_\mu \times V_\rho ] $ & I & 0 & $+0.765$(0) & 0 \\ 
$ [ V_\mu \times V_\nu ]
	[ V_\mu \times V_\rho ] $ & II & 0 & $-0.255$(0) & 0 \\ 
\hline
$ [ V_\mu \times A_\mu ]
	[ V_\mu \times A_\mu ] $ & I & 0 & $+0.171$(0) & 0 \\ 
$ [ V_\mu \times A_\mu ]
	[ V_\mu \times A_\mu ] $ & II & 0 & $-0.057$(0) & 0 \\ 
$ [ V_\mu \times A_\nu ]
	[ V_\mu \times A_\nu ] $ & I & 0 & $+0.224$(0) & 0 \\ 
$ [ V_\mu \times A_\nu ]
	[ V_\mu \times A_\nu ] $ & II & 0 & $-0.075$(0) & 0 \\ 
$ [ V_\mu \times A_\mu ]
	[ V_\mu \times A_\nu ] $ & I & 0 & $+0.048$(0) & 0 \\ 
$ [ V_\mu \times A_\mu ]
	[ V_\mu \times A_\nu ] $ & II & 0 & $-0.016$(0) & 0 \\ 
$ [ V_\mu \times A_\nu ]
	[ V_\mu \times A_\mu ] $ & I & 0 & $+0.048$(0) & 0 \\ 
$ [ V_\mu \times A_\nu ]
	[ V_\mu \times A_\mu ] $ & II & 0 & $-0.016$(0) & 0 \\ 
\hline
\end{tabular}
\end{center}
\caption{Renormalization constants of a color two trace four-fermion
operator at one loop: $({\cal O}^{Latt}_5)_{II}$ defined in Eq.
(\ref{eq:latt-basis}).
%
%
The $\hat{\Gamma}_{ij}$ matrix represents an anomalous dimension
defined in Eq.~(\ref{eq:ff-latt-Z}) and its values are given in
Eq.~(\ref{eq:ff-anomalous}).
The $\hat{C}^{Latt}_{ij}$ matrix is defined in
Eq.~(\ref{eq:ff-latt-Z}).
The $N_{ij}$ matrix represents the tadpole-improvement
defined in Eq.~(\ref{eq:ff-TI-N}).
$c_{TI} \equiv (4\pi)^2 / 12$.
All the greek indices are summed under the condition of
$\mu \ne \nu \ne \rho \ne \eta$.
}
\label{tab:ff-op-T-S-P-II-0}
\end{table}
\begin{table}[h]
\begin{center}
\begin{tabular}{ | l | c || c | r | r | }
\hline 
${\cal O}^{Latt(0)}_j $ 	& color trace 	& $\hat{\Gamma}_{ij}$
 	& $\hat{C}^{Latt}_{ij}$ 	& $ N_{ij} $ \\ \hline
\hline
$ [ T_{\mu\nu} \times S ]
	[ T_{\mu\nu} \times S ] $ & I & 0 & $+0.095$(0) & 0 \\ 
$ [ T_{\mu\nu} \times S ]
	[ T_{\mu\nu} \times S ] $ & II & 0 & $-0.032$(0) & 0 \\ 
$ [ T_{\mu\nu} \times T_{\mu\nu} ]
	[ T_{\mu\nu} \times T_{\mu\nu} ] $ & I & 0 & $-1.335$(0) & 0 \\ 
$ [ T_{\mu\nu} \times T_{\mu\nu} ]
	[ T_{\mu\nu} \times T_{\mu\nu} ] $ & II & 0 & $+0.445$(0) & 0 \\ 
$ [ T_{\mu\nu} \times T_{\mu\rho} ]
	[ T_{\mu\nu} \times T_{\mu\rho} ] $ & I & 0 & $-1.711$(0) & 0 \\ 
$ [ T_{\mu\nu} \times T_{\mu\rho} ]
	[ T_{\mu\nu} \times T_{\mu\rho} ] $ & II & 0 & $+0.570$(0) & 0 \\ 
$ [ T_{\mu\nu} \times T_{\mu\nu} ]
	[ T_{\mu\nu} \times T_{\mu\rho} ] $ & I & 0 & $-0.201$(0) & 0 \\ 
$ [ T_{\mu\nu} \times T_{\mu\nu} ]
	[ T_{\mu\nu} \times T_{\mu\rho} ] $ & II & 0 & $+0.067$(0) & 0 \\ 
$ [ T_{\mu\nu} \times T_{\mu\rho} ]
	[ T_{\mu\nu} \times T_{\mu\nu} ] $ & I & 0 & $-0.201$(0) & 0 \\ 
$ [ T_{\mu\nu} \times T_{\mu\rho} ]
	[ T_{\mu\nu} \times T_{\mu\nu} ] $ & II & 0 & $+0.067$(0) & 0 \\ 
$ [ T_{\mu\nu} \times T_{\mu\rho} ]
	[ T_{\mu\nu} \times T_{\nu\rho} ] $ & I & 0 & $-0.445$(0) & 0 \\ 
$ [ T_{\mu\nu} \times T_{\mu\rho} ]
	[ T_{\mu\nu} \times T_{\nu\rho} ] $ & II & 0 & $+0.148$(0) & 0 \\ 
$ [ T_{\mu\nu} \times T_{\mu\rho} ]
	[ T_{\mu\nu} \times T_{\mu\eta} ] $ & I & 0 & $-0.042$(0) & 0 \\ 
$ [ T_{\mu\nu} \times T_{\mu\rho} ]
	[ T_{\mu\nu} \times T_{\mu\eta} ] $ & II & 0 & $+0.014$(0) & 0 \\ 
$ [ T_{\mu\nu} \times T_{\rho\eta} ]
	[ T_{\mu\nu} \times T_{\rho\eta} ] $ & I & 0 & $-1.335$(0) & 0 \\ 
$ [ T_{\mu\nu} \times T_{\rho\eta} ]
	[ T_{\mu\nu} \times T_{\rho\eta} ] $ & II & 0 & $+0.445$(0) & 0 \\ 
$ [ T_{\mu\nu} \times P ]
	[ T_{\mu\nu} \times P ] $ & I & $+2$ & $+1.500$(0) & 0 \\ 
$ [ T_{\mu\nu} \times P ]
	[ T_{\mu\nu} \times P ] $ & II & $-10/3$ & $-15.119$(1) & $+c_{TI}$ \\ 
\hline
\end{tabular}
\end{center}
\caption{Renormalization constants of a color two trace four-fermion
operator at one loop, $({\cal O}^{Latt}_5)_{II}$ defined in Eq. 
(\ref{eq:latt-basis}).
%
%
The $\hat{\Gamma}_{ij}$ matrix represents an anomalous dimension
defined in Eq.~(\ref{eq:ff-latt-Z}) and its values are given in
Eq.~(\ref{eq:ff-anomalous}).
The $\hat{C}^{Latt}_{ij}$ matrix is defined in
Eq.~(\ref{eq:ff-latt-Z}).
The $N_{ij}$ matrix represents the tadpole-improvement
defined in Eq.~(\ref{eq:ff-TI-N}).
$c_{TI} \equiv (4\pi)^2 / 12$.
All the greek indices are summed under the condition of
$\mu \ne \nu \ne \rho \ne \eta$.
}
\label{tab:ff-op-T-S-P-II-1}
\end{table}
\begin{table}[h]
\begin{center}
\begin{tabular}{ | l | c || c | r | r | }
\hline 
${\cal O}^{Latt(0)}_j $ 	& color trace 	& $\hat{\Gamma}_{ij}$
 	& $\hat{C}^{Latt}_{ij}$ 	& $ N_{ij} $ \\ \hline
\hline
$ [ A_\mu \times V_\mu ]
	[ A_\mu \times V_\mu ] $ & I & 0 & $+1.504$(0) & 0 \\ 
$ [ A_\mu \times V_\mu ]
	[ A_\mu \times V_\mu ] $ & II & 0 & $-0.501$(0) & 0 \\ 
$ [ A_\mu \times V_\nu ]
	[ A_\mu \times V_\nu ] $ & I & 0 & $+0.955$(0) & 0 \\ 
$ [ A_\mu \times V_\nu ]
	[ A_\mu \times V_\nu ] $ & II & 0 & $-0.318$(0) & 0 \\ 
$ [ A_\mu \times V_\mu ]
	[ A_\mu \times V_\nu ] $ & I & 0 & $-0.382$(0) & 0 \\ 
$ [ A_\mu \times V_\mu ]
	[ A_\mu \times V_\nu ] $ & II & 0 & $+0.127$(0) & 0 \\ 
$ [ A_\mu \times V_\nu ]
	[ A_\mu \times V_\mu ] $ & I & 0 & $-0.382$(0) & 0 \\ 
$ [ A_\mu \times V_\nu ]
	[ A_\mu \times V_\mu ] $ & II & 0 & $+0.127$(0) & 0 \\ 
\hline
$ [ A_\mu \times A_\mu ]
	[ A_\mu \times A_\mu ] $ & I & 0 & $-0.673$(0) & 0 \\ 
$ [ A_\mu \times A_\mu ]
	[ A_\mu \times A_\mu ] $ & II & 0 & $+0.224$(0) & 0 \\ 
$ [ A_\mu \times A_\nu ]
	[ A_\mu \times A_\nu ] $ & I & 0 & $-1.014$(0) & 0 \\ 
$ [ A_\mu \times A_\nu ]
	[ A_\mu \times A_\nu ] $ & II & 0 & $+0.338$(0) & 0 \\ 
$ [ A_\mu \times A_\mu ]
	[ A_\mu \times A_\nu ] $ & I & 0 & $-0.143$(0) & 0 \\ 
$ [ A_\mu \times A_\mu ]
	[ A_\mu \times A_\nu ] $ & II & 0 & $+0.048$(0) & 0 \\ 
$ [ A_\mu \times A_\nu ]
	[ A_\mu \times A_\mu ] $ & I & 0 & $-0.143$(0) & 0 \\ 
$ [ A_\mu \times A_\nu ]
	[ A_\mu \times A_\mu ] $ & II & 0 & $+0.048$(0) & 0 \\ 
\hline
$ [ P \times S ]
	[ P \times S ] $ & I & 0 & $-0.286$(0) & 0 \\ 
$ [ P \times S ]
	[ P \times S ] $ & II & 0 & $+0.095$(0) & 0 \\ 
$ [ P \times T_{\mu\nu} ]
	[ P \times T_{\mu\nu} ] $ & I & 0 & $+0.419$(0) & 0 \\ 
$ [ P \times T_{\mu\nu} ]
	[ P \times T_{\mu\nu} ] $ & II & 0 & $-0.140$(0) & 0 \\ 
$ [ P \times T_{\mu\nu} ]
	[ P \times T_{\mu\rho} ] $ & I & 0 & $+0.042$(0) & 0 \\ 
$ [ P \times T_{\mu\nu} ]
	[ P \times T_{\mu\rho} ] $ & II & 0 & $-0.014$(0) & 0 \\ 
$ [ P \times P ]
	[ P \times P ] $ & I & $-6$ & $-4.500$(0) & 0 \\ 
$ [ P \times P ]
	[ P \times P ] $ & II & $-6$ & $-54.128$(1) & $+c_{TI}$ \\ 
\hline
\end{tabular}
\end{center}
\caption{Renormalization constants of a color two trace four-fermion
operator at one loop, $({\cal O}^{Latt}_5)_{II}$ defined in Eq. 
(\ref{eq:latt-basis}).
%
%
The $\hat{\Gamma}_{ij}$ matrix represents an anomalous dimension
defined in Eq.~(\ref{eq:ff-latt-Z}) and its values are given in
Eq.~(\ref{eq:ff-anomalous}).
The $\hat{C}^{Latt}_{ij}$ matrix is defined in
Eq.~(\ref{eq:ff-latt-Z}).
The $N_{ij}$ matrix represents the tadpole-improvement
defined in Eq.~(\ref{eq:ff-TI-N}).
$c_{TI} \equiv (4\pi)^2 / 12$.
All the greek indices are summed under the condition of
$\mu \ne \nu \ne \rho \ne \eta$.
}
\label{tab:ff-op-T-S-P-II-2}
\end{table}

\end{document}